\documentclass[usenatbib,useAMS]{mn2e}
\usepackage{epsf}
\usepackage{epsfig}
\usepackage{amsmath,amssymb}
\usepackage{times}
\usepackage{natbib}
\usepackage{dcolumn}
\RequirePackage{amsmath}
\RequirePackage{amssymb}        
\RequirePackage{wasysym}

\title{Stellar Winds and Mass Loss from Extreme Helium Stars}

\author[C.S. Jeffery \& W.-R. Hamann ]
       {
C.S. Jeffery\thanks{E-mail: csj@arm.ac.uk} 
\& 
W.-R. Hamann\thanks{E-mail: wrh@astro.physik.uni-potsdam.de}\\
Armagh Observatory, College Hill, Armagh BT61 9DG, Northern Ireland\\
Institut f\"ur Physik und Astronomie, Universit\"at Potsdam, University Campus Golm, Haus 28, Karl-Liebknecht-Str. 24/25, 14476 Potsdam
}

\date{Accepted .....
      Received ..... ;
      in original form .....}

\pagerange{\pageref{firstpage}--\pageref{lastpage}}
\pubyear{2008}

\begin{document}

\maketitle

\label{firstpage}

\begin{abstract}
Extreme helium stars are very rare low-mass supergiants in a 
late stage of evolution. They are probably contracting to become white dwarfs 
following a violent phase of evolution which caused them to become
hydrogen-deficient giants, possibly R\,CrB stars. 
Using the latest generation of models for spherically
expanding stellar atmospheres, we set out to measure mass-loss rates
for a representative fraction of these stars. We have used 
high-resolution ultraviolet and optical spectra, and
ultraviolet, optical and near-infrared photometry from a variety of
archives. Overall atmospheric parameters have mostly been taken from
previous analyses and checked for consistency. Mass-loss rates were 
measured by fitting the P-Cygni and asymmetric profiles of C, N and Si
ultraviolet resonance lines and lie in the range $10^{-10} -
10^{-7} {\rm M_{\odot} yr^{-1}}$. These rates follow a
Castor-type ($\dot{M} \propto L^{1.5}$) relation 
marking a lower limit for the mass loss from hot stars of all kinds.
The mass-loss rates of the studied stars also show a strong
correlation with their proximity to the Eddington limit. 
There is no firm evidence for variability in the stellar wind, 
although photospheric pulsations have been reported in many cases. 
\noindent
\end{abstract}


\section{Introduction}              
\label{intro}

\noindent 

Amongst the more unusual groups of stars in the Galaxy is a collection 
of low-mass but nevertheless luminous stars from whose surfaces hydrogen
is almost completely absent. Instead they are composed principally of
helium and some 3\% carbon (by mass).  The collection includes R\,CrB
stars, with rough spectral classes F and G, extreme helium stars, with
classes A and B, and some helium subdwarf O stars \citep{jeffery08b}. It
had long been questioned whether these stars are hydrogen-deficient
because their outer layers have been stripped by a stellar wind, as in
the massive Wolf-Rayet stars, or diluted by deep mixing of the stellar
envelope, as in the low-mass Wolf-Rayet central stars of planetary
nebulae, or by some more exotic process \citep{schoenberner86}.
Increasing evidence has accrued that these stars are the result of a
merger between a helium- and a carbon/oxygen white dwarf.  In such stars
the helium white dwarf is entirely disrupted, wrapping itself around its
more massive companion, whereupon a helium shell is ignited and the star
becomes a helium supergiant for a few tens of thousands of years, before
contracting to become a massive C/O white dwarf
\citep{webbink84,iben84,saio02,pandey06,clayton07}. While the arguments
for the white dwarf merger model are compelling,  sufficient questions
remain that it is important to extract as much physical information
concerning these stars as possible. 

It has been known since the early days of the International Ultraviolet
Explorer (IUE) \citep{kondo89} that some of the hotter (B- and O-type)
hydrogen-deficient stars showed significant P\,Cygni-type line profiles
in strong ultraviolet resonance lines, indicating the presence of a
stellar wind and hence mass loss from the stellar surface
\citep{darius79, giddings81,hamann82, jeffery86, dudley92}.  Estimates
of mass-loss rates of between $10^{-11}$ and $10^{-7} {\rm M_{\odot}
yr^{-1}}$ from these early studies showed that the current stellar wind
is unlikely to be responsible for stripping the entire hydrogen
envelope.

However, the study of the winds in these stars is important for several
other reasons. First, it is well known that a relation exists between
luminosity and mass-loss rate (or, more precisely, wind momentum)
\citep{kudritzki99}, so that the latter becomes a proxy  for the former.
This is extremely useful when distances are not known. Moreover, models
for pre-white dwarf contraction give a direct relation between
luminosity and mass \citep{saio88b,saio02}.

The drawback is that the modelling of a stellar wind requires a
difficult computation in order to satisfy conditions of radiative and
statistical equilibrium, and momentum and energy conservation.
Specifically, the conditions of local thermodynamic and hydrostatic
equilibrium no longer hold, and it is necessary to compute a
self-consistent multi-level multi-atomic model in an expanding medium.
Fortunately, in the quarter century since this topic was last addressed,
our ability to perform such calculations has improved by orders of
magnitude, so that a re-examination of the question is timely.
Additional data from IUE, the Hubble Space Telescope (HST), the Far
Ultraviolet Space Explorer (FUSE), the T\"ubingen Echelle Spectrograph
aboard the Orbiting and Retrievable Far and Extreme Ultraviolet
Spectrometer (ORFEUS/TUES) and several ground-based telescopes provide
an additional opportunity to improve on earlier results. In the cases of
the hottest (O-type) helium stars, the LTE assumption is of little
value, so that a study of this type provides an opportunity to gain new
knowledge about the surfaces of these stars.

This paper addresses the following questions. Do early-type
low-mass helium stars show conventional radiatively driven stellar
winds and, if so, what are the associated mass-loss rates? Can such
a mass-loss rate be used as a proxy for the luminosity, and hence
mass, of the star in question? Are the results consistent with the 
predictions of stellar evolution theory? 

Section 2 describes the
observational data available for such a study and explains which 
data have been used and why. It gives more information
about the particular stars examined -- namely 
BD$+37^{\circ}442$, BD$+37^{\circ}1977$, HD160641, BD$-9^{\circ}4395$, 
BD$+10^{\circ}2179$, and HD144941. Section 3 describes the theoretical
framework for the wind analysis, including the self-consistent wind 
models, the formal solutions and the methods used to fit the
observations. Section 4 presents the measurements for individual
stars, and Section 5 concludes with an assessment and interpretation of the 
measurements.

\section{Observational Data}

\noindent

To determine stellar wind properties it is necessary to  observe the
profiles of various spectral lines. Most of the lines of interest lie in
the ultraviolet and far ultraviolet; therefore high-resolution
spectroscopy from space is required. To support these data, the surface
temperature and chemical composition of the star also need to be known.
The former is best described by fitting the overall flux distribution,
the latter by fine analysis of the high-resolution photospheric
spectrum. For three of the stars studied here, measurements of both have
already been made; these show some discrepancies from one publication to
another,  which can be attributed to improvements in either the
observations or the model atmospheres. The following subsections
describe the stars selected for study, the ultraviolet data available
for wind analysis, and other supporting ground-based spectroscopy and
photometry.

\subsection{Sample Selection}

\begin{figure*}
\includegraphics[width=12cm,angle=0]{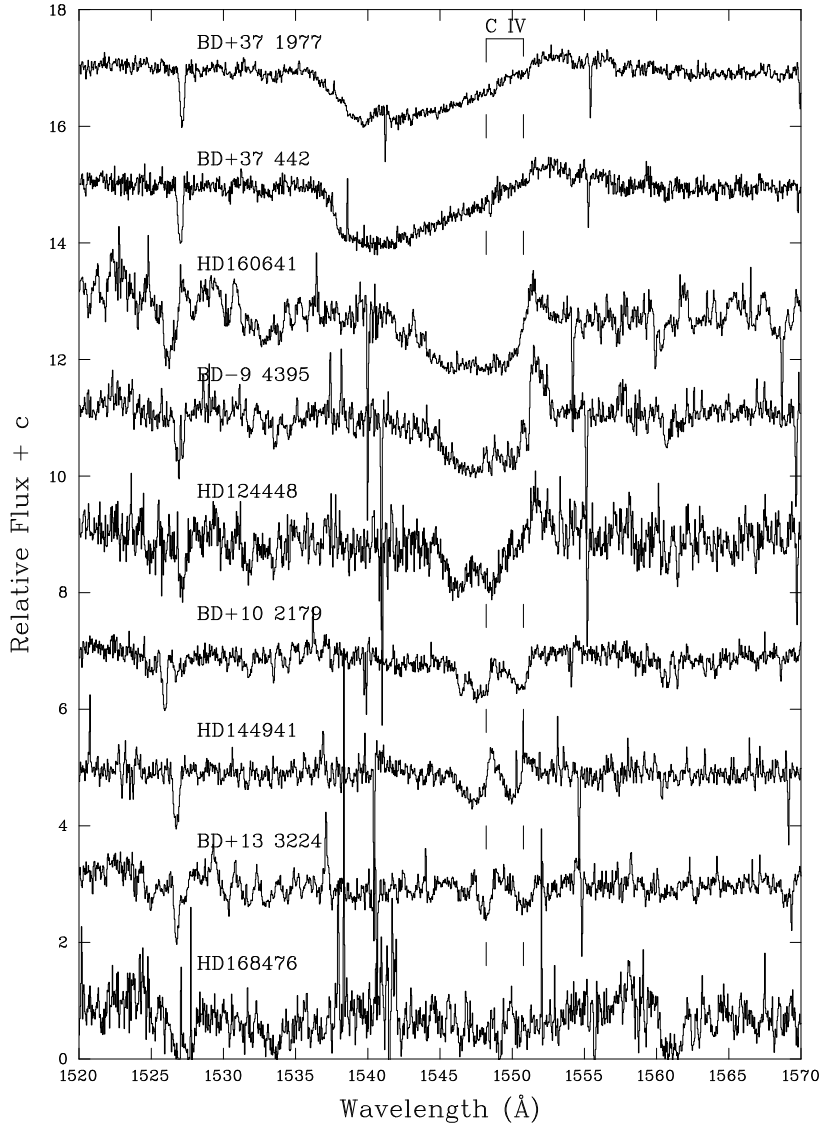}
\caption[C IV]{The C{\sc iv} ultraviolet resonance doublet in all extreme
helium stars observed with IUE. The rest position of the doublet is
marked. The stars are arranged in decreasing effective temperature from
top to bottom. All spectra are from SWP HIRES images and are the result
of merging the available spectra for each star (Table~\ref{t_data}).
They have been wavelength-corrected to bring photospheric lines close to
laboratory wavelengths. The strong feature around 1526\AA\ is due to
interstellar absorption, the sharp features around 1540, 1555 and
1569\AA\ are due to fiducial marks on the camera.}
\label{f_civ}
\end{figure*}

The primary criterion for selection is that targets should be bright
enough for high-resolution ultraviolet spectroscopy, and so  must be hot
with low interstellar extinction. This limited the original study of
winds in extreme helium stars \citep{hamann82} to HD160641,
BD$-9^{\circ}4395$,  and BD$+10^{\circ}2179$. Other extreme helium stars
with effective temperature\footnote{We distinguish the effective
temperature $T_{\rm eff}$ obtained from analyses of  photospheric
spectra from the surface temperature $T_{\ast}$ conventionally used in
stellar wind models (\S\,3).} $T_{\rm eff} > 15\,000\,{\rm K}$ include
BD$+13^{\circ}3224$, LSS\,3184,  LS\,IV$+6^{\circ}2$, LSS\,5121,
DY\,Cen, LSS\,4357, LSE\,78, LSS\,99, LS\,IV$-14\,109$,  HD124448 and
MV\,Sgr \citep{jeffery01a}. While all have been observed with IUE, none
show evidence of strong P\,Cygni line profiles at low resolution. The
hottest stars (LS\,IV$+6^{\circ}2$ and LSS\,5121)  are heavily reddened
and have not yet been observed at high resolution.  Thus the secondary
selection criterion is that suitable data are already
available. Figure~\ref{f_civ} shows a part of the spectrum of those
stars which satisfy these criteria in the vicinity of 
the C{\sc iv} resonance double at 1548-50\AA. Strong P\,Cygni profiles
are evident in the hottest stars (top), weakening towards lower effective
temperature (bottom). For BD$+10^{\circ}2179$ and HD144941, the  C{\sc iv}
lines are asymmetric. For BD$+13^{\circ}3224$, they are slightly
broadened and possibly purely photospheric. For HD168476, only the
interstellar component is readily identified.

In addition to these classical extreme helium stars we included three
closely-related stars.  
HD144941 has a hydrogen abundance around 3\% by mass
\citep{harrison97}, and is carbon weak. It has been linked to
low-luminosity extreme helium stars like BD$+13^{\circ}3224$, although
the nature of the connection is unclear. The UV spectrum shows
evidence of a stellar wind \citep{jeffery86}.  

It is not known whether the two helium sdO stars
BD$+37^{\circ}442$ and BD$+37^{\circ}1977$  represent very hot extreme helium
stars, although they are known to be hydrogen-deficient
\citep{wolff74} 
and very luminous relative to other helium-rich sdO stars.  
No complete quantitative analysis has been formally published for either star. 
A non-LTE study of the winds in both stars gives mass-loss rates \citep{giddings81}
and was reported by \citet{darius79}. 
A photospheric analysis of BD$+37^{\circ}442$ by Husfeld was
reported by \citet{bauer95} and shows   
H and C abundances similar to other 
extreme helium stars.  

Finally, we omitted
BD$+13^{\circ}3224$, HD124448 and HD168476 from our analysis on 
the grounds that these data either showed little evidence for a 
stellar wind, or were too noisy.

%

\subsection{Ultraviolet Spectroscopy}

High-resolution ultraviolet spectra of extreme helium stars have been
obtained with four spacecraft, namely the International Ultraviolet
Explorer (IUE), 
the Hubble Space Telescope (HST), The Far Ultraviolet Explorer (FUSE)
and the Orbiting and Retrievable Far and Extreme Ultraviolet
Spectrometer (ORFEUS II).  

A summary of the data available for extreme helium stars with $T_{\rm
  eff}>12\,000$K is given in Table \ref{t_data}. In four cases, spectra were 
obtained in overlapping wavelength intervals by more than one
instrument, and found to agree well. Where multiple exposures of
a given object were obtained with the same instrument, the reduced
spectra were combined to increase the overall signal-to-noise
ratio, using the individual exposure times as weights.

The greatest body of ultraviolet spectroscopy for extreme helium stars
continues to be SWP HIRES images obtained with IUE. Despite the larger
collecting  area of the Hubble Space Telescope, shorter exposure times
mean that STIS images which include the key wind lines (C\,{\sc iii},
C\,{\sc iv}, Si\,{\sc iv}, N\,{\sc v}) have lower  signal-to-noise than the
corresponding merged IUE data. ORFEUS data for BD$+37^{\circ}1977$ are
of considerably higher S/N than the merged IUE spectrum, so these data
have been used at wavelengths shortward of 1400\AA. For
BD$+37^{\circ}442$, the IUE and ORFEUS data are of comparable quality.

\begin{table*}
\caption{High-resolution spectroscopic data used in the current analysis}
\label{t_data}
\begin{tabular}{lcclrcc}
\hline 
Star \rule[0mm]{0mm}{3.25mm} & Instrument (Observer) & Image Number & Date & $t_{\rm exp}$ [s] 
& $\lambda$ [\AA] & Res. [\AA]\\
\hline 

BD$+37^{\circ}1977$ \rule[0mm]{0mm}{3.25mm}
 & IUE SWP HL & 06766 & 1979-10-05  & 3600 & 1150 -- 1980 \\
 & IUE SWP HL & 07248 & 1979-11-28  & 2643 & 1150 -- 1980 \\
 & MacDonald 2.7m Coude  & &        &      & 3700 -- 7000 & 0.15 \\[1mm]

BD$+37^{\circ}442$ 
 & IUE SWP HL  & 06768    & 1979-10-05 & 3600 & 1150 -- 1980 \\
 & ORFEUS TUES & 5209\_3  & 1996-11-30 & 2341 &  904 -- 1410 \\
 & Calar Alto 2.2m FOCES  & &      &      & 3980 -- 6600 & 0.1 \\[1mm]

HD\,160641 
 & HST STIS 0.2X0.2 E140M & O66V02010 & 2000-10-23 & 150 & 1145 -- 1729\\
 & ESO 3.6m CASPEC (1)    &   mean    & 1984-04-03, 1985-04-09 &     & 3900 -- 5160 & 0.15\\
 & AAT 3.9m UCLES  (2)    &   mean    & 2005-08    &      & 3900 -- 5160 & 0.15
 \\[1mm]

BD$-9^{\circ}4395$ 
 & IUE SWP HL & 13826 & 1981-04-28 & 23880 & 1150 -- 1980 \\
 & IUE SWP HL & 30797 & 1987-04-17 & 24240 & 1150 -- 1980 \\
 & IUE SWP HL & 30808 & 1987-04-19 & 23520 & 1150 -- 1980 \\
 & IUE SWP HL & 30814 & 1987-04-20 & 20460 & 1150 -- 1980 \\
 & IUE SWP HL & 30818 & 1987-04-21 & 21540 & 1150 -- 1980 \\
 & ESO 3.6m CASPEC (3)& mean & 1987-04    &       & 3994 -- 4988 & 0.15 \\
 & ESO 3.6m CASPEC (3)& mean & 1987-04    &       & 5778 -- 6790 & 0.15 \\[1mm] 

BD$+10^{\circ}2179$ 
 & HST STIS 0.2X0.2 E140M & O66V04010 & 2001-02-11 & 144 & 1145 -- 1729 \\
 & ESO 3.6m CASPEC (1)    &           & 1985-04-08  &  & 3790 -- 4810 & 0.15 \\[1mm]

HD\,144941 
 & IUE SWP HL & 23961 & 1984-09-14 & 12600 & 1150 -- 1980 \\
 & IUE SWP HL & 23962 & 1984-09-14 &  8940 & 1150 -- 1980 \\
 & AAT 3.9m UCLES (4) &   & 1995-04-15  &    & 3900 -- 5160 & 0.15 \\
\hline 
\end{tabular}\vspace{1mm}

\parbox{\textwidth}{\centering
Observer: 
(1) Heber; 
(2) Ahmad, \,{S}ah\'{i}n \& Jeffery;  
(3) Heber \& Jeffery;
(4) Harrison \& Jeffery}

\end{table*}

\subsection{Optical Spectroscopy}

Blue-optical spectra have been used to ensure that the wind  models are
consistent with the global emergent spectrum and, in particular, with
the profiles of strong lines formed in the photosphere and unaffected by
the wind.  These data come from a variety  of sources
(Table~\ref{t_data}) and are also used with caution. Three of
the targets are known or suspected to be variable; the adopted optical
spectra  of HD160641 and  BD$-9^{\circ}4395$ are direct averages of
time-series spectra obtained over four and three nights respectively.
Both stars are pulsating and and show substantial line-profile
variability on a timescale of hours \citep{jeffery92,wright06}. 
Variability with $P\sim 590 {\rm s}$ in BD$+37^{\circ}442$
\citep{bartolini82} has not been confirmed.

\subsection{ Ultraviolet, Optical and Infrared Spectrophotometry}

Finally, to ensure conservation of total flux and to check the
interstellar extinction, spectrophotometry  from the ultraviolet to
infrared was collated. UV spectrophotometry was the same as used by
\citet{jeffery01a}, with the addition of IUE  images SWP05642, LWR04892
and LWR02805 for BD$+37^{\circ}442$.  Optical photometry (Johnson $B$ and
$V$) were taken from SIMBAD,  and infrared fluxes were retrieved from 2MASS.

\section{Theoretical Framework}
\label{Theory}

For analyzing the stars, we compare the observations with synthetic
spectra calculated with the Potsdam Wolf-Rayet (PoWR) model atmospheres
code \citep[and references therein]{PoWR}. 
This code solves the non-LTE radiative transfer in a 
spherically symmetric expanding atmosphere. Mass-loss rate and wind 
velocity are free parameters of the models.  

Wind inhomogeneities (``clumping'') are accounted for in a first-order
approximation, assuming that optically thin clumps fill a volume
fraction $f_{\rm V}$ while the interclump space is void. Thus the matter
density in the clumps is higher by a factor $D = f_{\rm V}^{-1}$,
compared to an un-clumped model with the same parameters. $D$\,=\,4 is
arbitrarily assumed throughout this paper. Nothing is known about the
wind homogeneities in our program stars. Even for the winds from massive
stars, the clumping contrast is highly debated
\citep[see][]{clumpingworkshop}. Fortunately, this issue has no impact
on the present study, since we derive the mass-loss rates from resonance
lines. Since their (mainly line-scattering) opacity scales linearly with
density, inhomogeneity does not change the average opacity as long as
the individual clumps are optically thin \citep{macroclumping}. Moreover,
since the resonance lines are not strongly saturated in the observed
spectra our program stars, the clumps are certainly not optically thick.

Detailed ions of H, He, C\,{\sc ii} -- C\,{\sc iv}, N\,{\sc ii} --
N\,{\sc v} and Si\,{\sc iii} -- Si\,{\sc iv}  are taken into account in
our models. About 350 non-LTE levels and 4500 line transitions are
explicitely treated. Because of the unsettled questions about how to
include dielectronic recombination properly, this process is neglected.

\begin{figure*}
\includegraphics[width=\textwidth,angle=0]{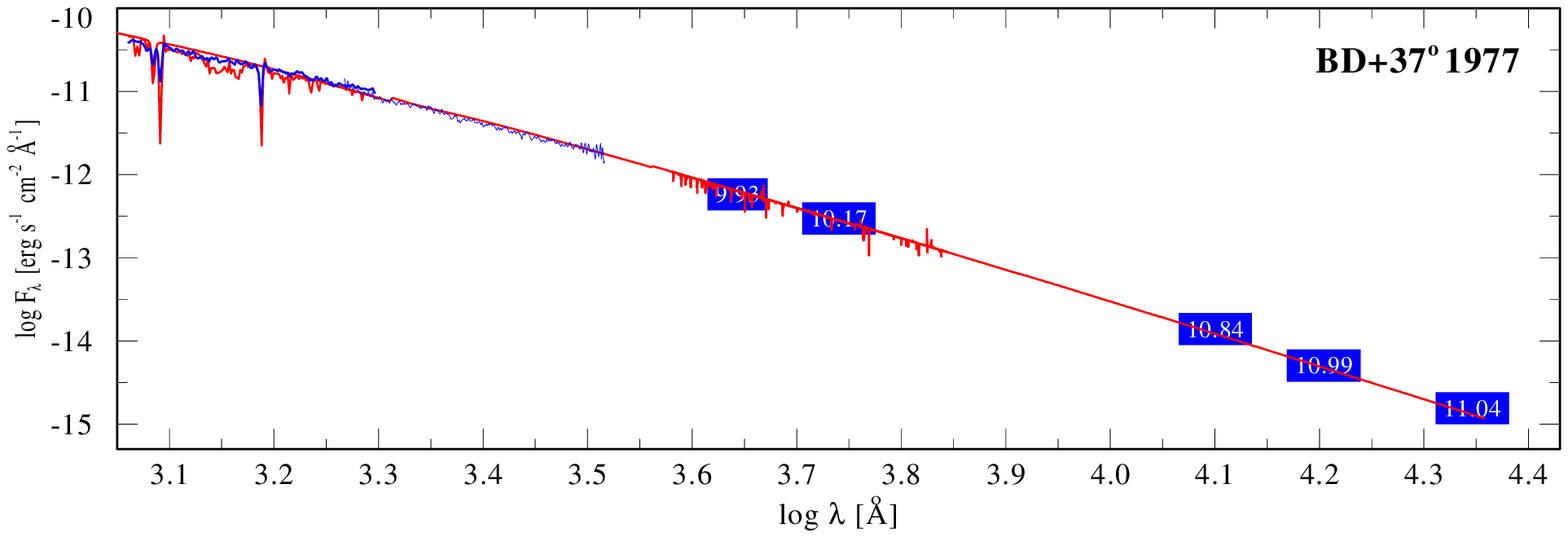}
\caption[1977A]{Spectral energy distribution (SED) for BD$+37{^\circ}1977$
from the UV to the near IR. The observations comprise 
low-resolution IUE data (thin blue line), visual photometry (blue blocks 
with magnitudes) and near-IR photometry from 2MASS. The model continuum
(see Table\,\ref{t_phys_new} for the parameters) is shown 
by the straight red line, in parts of the spectrum augmented by the 
synthetic line spectrum. Since the model luminosity was adopted from 
previous work, the SED fit is achieved by adjusting the color excess 
$E_{B-V}$ and the distance modulus $D\!M$.}
\label{f_sed_BD+37D1977}
\end{figure*}

The models account for line blanketing by iron and other
iron-group elements. About $10^5$ energy levels and $10^7$ line
transitions between those levels are taken into account in the
approximation of the ``superlevel'' approach.

As already mentioned, the velocity field must be pre-specified. In the
subsonic region, $v(r)$ is defined such that a hydrostatic density
stratification is approached. For the supersonic part we adopt the usual
$\beta$-law, basically of the form $v(r) = v_\infty (1 - 1/r)^\beta$,
with the terminal velocity $v_\infty$ being a free parameter. The
exponent $\beta$ is chosen between 0.7 and 1.0 such that the wind
profiles are reproduced best. For some program stars we employ a
``double beta law'', which means that a second term of the same
$\beta$-law form is added to the velocity, but with a large $\beta=4$.
Typically, 40\% of the terminal velocity is attributed to this second
$\beta$ term. Its effect is that the velocity $v(r)$ does not approach
the terminal velocity too soon. The ``double beta law'' makes the blue
edges of P-Cygni profiles less steep. This often fits the
observation better, and has also found theoretical support from
hydrodynamically consistent models \citep{wr111}.

\begin{table*}
\caption{Previously published physical parameters for the program 
stars. 
 Abundances $X$ are mass fraction in per cent.}
\label{t_phys_old}
\begin{tabular}{r cccccc r ccccc r}
\hline 
\multicolumn{5}{l}{Star}\\
& $T_{\rm eff}$ & $E_{B-V}$ & $\log g$ & $\log L$ &
     $v_{\infty}$ & 
     $\log \dot{M}$ & Ref. 
       & $X_{\rm H}$ & $X_{\rm C}$ & $X_{\rm N}$ & $X_{\rm Si}$ &
       $X_{\rm Fe}$ &     Ref.  \\
& [kK] & [mag] & [${\rm cm\,s^{-2}}$] & [${\rm L_{\odot}}$] &
  [${\rm km\,s^{-1}}$] & 
  [${\rm M_{\odot} yr^{-1}}$] & 
  & [\%] & [\%] & [\%] & [\%] & [\%] &      \\[1mm]
\hline 
\multicolumn{5}{l}{BD$+37^{\circ}1977$ }\\
& 41.0 & 0.00 & & & & & 1 & \\
& 55.0 & 0.00 & 4.0 & 4.4 & 2200 & $-8.2 \ldots -8.0$ & 2 \\
&      &      &     &     &      &  $-9.8$  & 3 \\

\multicolumn{5}{l}{BD$+37^{\circ}442$ }\\
& 60.0 & 0.0  & 4.0 & 4.5 &      &                    & 4 & 
                                           0 & 2.5 & 0.3  & 0.08 &  & 5 \\
& 55.0 & 0.00 & 4.0 & 4.4 & 2200 & $-8.2 \ldots -8.0$ & 2 \\
&      &      &     &     &      &  $-9.4$  & 3 \\ 
 
\multicolumn{5}{l}{HD160641}\\ 
& 31.6 & 0.45 &     &     &     &                    & 1 & 
                                  0 & 0.3 & 0.5 & 0.06 &   & 6 \\
& 34   &      & 2.8 &     &     &                    & 7 \\
&      &      &     & 4.5 & 550 & $-8.2 \ldots -7.2$ & 8 \\
&      &      &     &     &     &  $-10.8$           & 3 \\

\multicolumn{5}{l}{BD$-9^{\circ}4395$}\\
& 22.7 & 0.24 & 2.55 & 4.1$^\dagger$& & & 1,9 & 
                                  0.04 & 1.3 & 0.09 & 0.14 & 0.017 & 9 \\
&      &      &     &    & 600 & $-8.5 \ldots -7.7$ & 8 \\
&      &      &     &     &      &  $-10.5$  & 3 \\

\multicolumn{5}{l}{BD$+10^{\circ}2179$}\\
& 19.5 & 0.00 & 2.6 & 3.7$^{\dagger}$ & & & 1,10 &
                                     0 & 3 & 0.13 & 0.04 & 0.01 & 10 \\
& 16.9 & 0.00 & 2.6 & 3.5$^{\dagger}$ &     &    &  11  &
                                 0 & 2.2 & 0.08 & 0.013 & 0.006 & 11 \\
&      &      &     &     & 400 & $-11.0 \ldots -8.9$ & 8 \\
&      &      &     &     &      &  $-14.4$  & 3 \\

\multicolumn{5}{l}{HD144941}\\
& 23.2 & 0.25 & 3.9  & 2.7$^{\dagger}$ &    &  & 12 & 
                            1.4 & 0.005 & 0.003 & 0.0025 & 0.002 & 12,13 \\
& 27.8 & 0.25 &     &    &     &                    & 1  \\
&      &      &     &                 & 350 & $-9.7 \ldots -9.0$ & 14  \\[1mm] 

\hline 

\end{tabular}
\begin{tabular}{rlll}
Notes: 
& \multicolumn{3}{l}{$\dagger$: estimated from $T_{\rm eff}$, $\log g$ 
using the $M_{\rm core}-L_{\rm shell}$ relation for He-shell burning
stars \citep{jeffery88}.} \\
References:
& 1: \citet{jeffery01a}
& 2: \citet{darius79,giddings81}
& 3. \citet{dudley92} \\

& 4: \citet{husfeld87}
& 5: \citet{bauer95} 
& 6: \citet{all54} \\

& 7: \citet{rauch96} 
& 8: \citet{hamann82}
& 9: \citet{jeffery92}\\

& 10: \citet{heber83}
& 11: \citet{pandey06} 
& 12: \citet{harrison97} \\

& 13: \citet{jeffery97}
& 14: \citet{jeffery86} \\

\end{tabular}
\end{table*}

Further input parameters to the PoWR models are the luminosity $L$ and the
``stellar temperature'' $T_*$. The latter refers, via the
Stefan-Boltzmann law, to the radius $R_*$ which is the inner boundary of
the model atmosphere, by definition located at a Rosseland optical depth
of 20. In contrast, the usual ``effective temperature'' $T_{\rm eff}$
refers to the radius where the Rosseland optical depth is 2/3. In thick
expanding atmospheres, e.g.\ from Wolf-Rayet stars, this can be
considerably larger than $R_*$. The winds of our program stars are
rather weak, and hence the difference between $T_*$ and $T_{\rm eff}$ can
be neglected here.

The Doppler velocity $v_{\rm D}$ reflects random motions on small scales
(``microturbulence''). In stellar winds $v_{\rm D}$ is usually found to
be of the order of 10\% of the terminal wind velocity. Unfortunately the
PoWR code does not allow to vary $v_{\rm D}$ with the depth in the
atmosphere. Therefore we apply a smaller $v_{\rm D}$ value in the 
``formal integral'' when calculating the photospheric absorption 
spectrum in the optical, and a larger value for the UV spectrum with the 
wind lines. The millions of iron-group lines are always calculated with 
$v_{\rm D} = 50\,{\rm km\,s^{-1}}$ for numerical reasons, which is 
certainly too high for the photosphere and therefore may lead to a 
slight overestimate of the iron-line forest in the UV. 

In addition to Doppler broadening from microturbulence, radiation 
damping and pressure broadening of spectral lines are taken into account
in the formal integral. This is a new feature of the PoWR code not yet 
documented in previous publications. The broadening of H\,{\sc i} and
He\,{\sc ii} lines uses tables from \cite{VCS} and 
\citet{sch89a,sch89b,sch89c} 
respectively. 
Voigt functions are used for lines  of neutral
helium, with calculation of the coefficients following  \cite{Griem}.
Excited lines from other elements  are broadened by the quadratic Stark
effect following the approximation  from \cite{Cowley}. 

Pressure broadening is missing in the code for lines that are not covered
by the mentioned sources. We have no pressure broadening included yet
for resonance lines, which is a noticeable deficiency in our spectral
fits. This somewhat incomplete treatment of line broadening is due to
the historical fact that the PoWR code was developed for stellar wind
spectra where pressure broadening is negligible.

\section{Spectral analyses}
\label{Results}

Since the winds from our program stars are weak, their spectra mainly
form in the nearly hydrostatic photosphere. Except for the strongest
lines which are affected by the stellar wind, static, plane-parallel
non-LTE models therefore provide a reasonable approximation. All our
program stars have been previously studied with static models in
more or less detail. We adopt the parameters from these previous
studies, at least as a first guess. This refers to the effective
temperature $T_{\rm eff}$, the surface gravity $\log g$, the stellar
luminosity $L$, and the chemical composition (see
Table\,\ref{t_phys_old}). We check the fit of the photospheric
absorption line spectra with our models, and keep the parameters from
previous work whenever possible, but also encounter the need to adjust
parameters in a few cases (see below).

\subsection{BD+37$^{\circ}$1977}

Our fit procedure with the PoWR models is decribed now by the example of 
the first program star, BD$+37^{\circ}1977$. The observed and computed spectral 
energy distribution (SED) is shown in Fig.\,\ref{f_sed_BD+37D1977}. Since the
model luminosity was adopted from previous work, the SED fit is achieved
by adjusting the color excess $E_{B-V}$ and the distance modulus
$D\!M$ to the values given in Table\,\ref{t_phys_new}. 

\begin{figure*}
\includegraphics[width=17cm,angle=0]{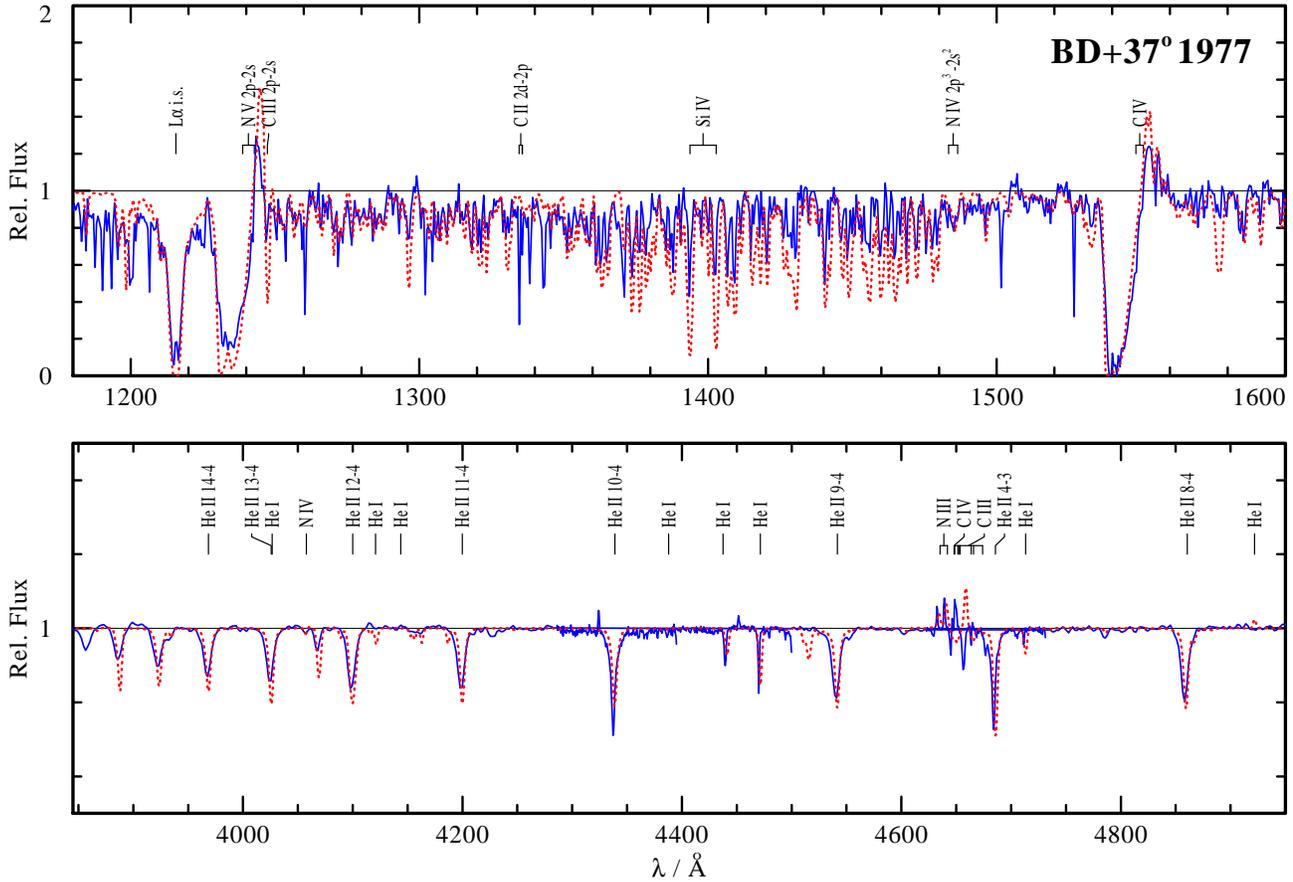}
\caption{Normalized line spectrum 
for BD$+37^{\circ}1977$ in the UV (top) and optical (bottom) range. 
The optical observation (blue line) has been normalized to the continuum 
``by eye'', while the absolutely calibrated UV spectrum was divided by 
the model continuum. The model spectrum (red dotted) is for the model 
parameters given in Table\,\ref{t_phys_new}. }
\label{f_lines_BD+37D1977}
\end{figure*}

The model SED depends of course also on the stellar parameters. Hence
we  have to look at the line spectrum as well, and the whole procedure
is in  fact iterated. The lower panel of Fig.\,\ref{f_lines_BD+37D1977}
shows  the optical spectrum, which is dominated by the lines from
neutral and  ionized helium. Getting the balance between the He\,{\sc i}
and the  He\,{\sc ii} lines right constrains the effective temperature.
Previous  (static) analyses give slightly discrepant values for $T_{\rm
eff}$  (41\,kK and 55\,kK, respectively, see Table\,\ref{t_phys_old}).
We find a good fit with $T_{\rm eff}  = 48\,{\rm kK}$. The
pressure-broadened  line wings are reasonably well reproduced with $\log
g = 4.0$ from the  previous work.  

Unfortunately, BD$+37^{\circ}1977$ was not analyzed yet for the chemical
abundances. However, BD$+37^{\circ}442$ can be considered as a
spectroscopic twin, and the composition of the latter has been studied
with static models by \cite{bauer95}. Therefore we adopt the abundances
from that paper. The iron mass fraction we set to half solar.
With this composition, the simulated ``iron forest'' in the UV still
appears a bit stronger than observed, as does the Si\,{\sc iv}
resonance doublet. But this problem can be attributed to the effect
of microturbulence. The UV range has been calculated with a
microturbulent velocity of 100\,km\,s$^{-1}$, which is adequate for the
P\,Cygni profiles from N\,{\sc v} and C\,{\sc iv} that are formed in
the very fast wind of this star. For the photospheric lines this value
is far too high, but the code allows only for a uniform value.
The optical range is calculated with a microturbulence of 
30\,km\,s$^{-1}$ which may be adequate for the photosphere.

\begin{figure}
\includegraphics[width=\columnwidth]{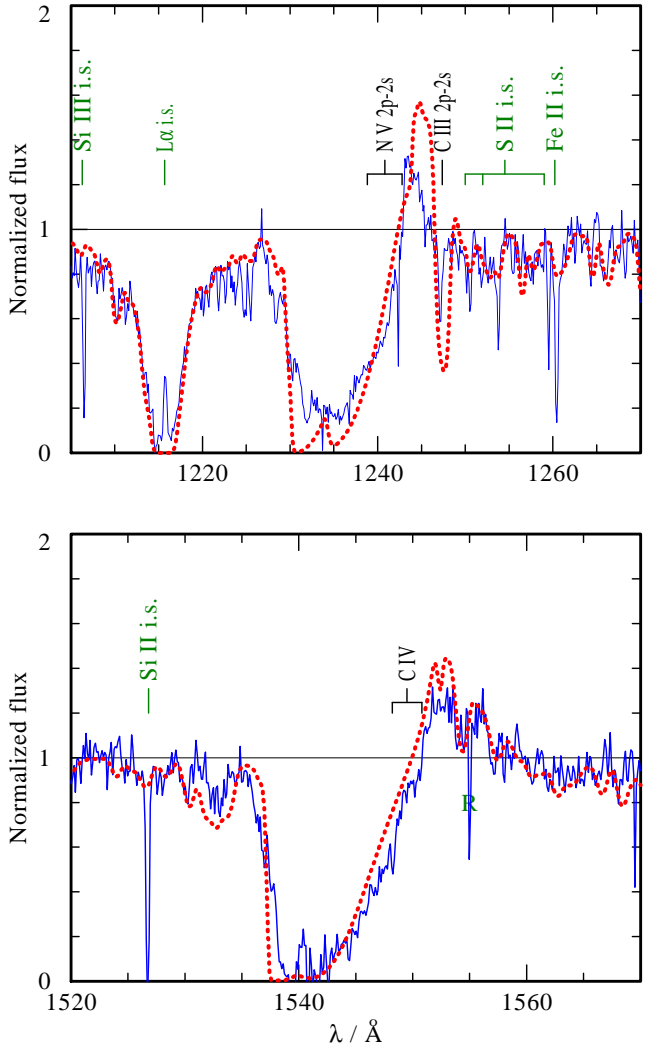}
\caption{Normalized line spectrum  for BD$+37^{\circ}1977$ as in
Fig.\,\ref{f_lines_BD+37D1977}, zoomed at the UV resonance doublets of
N\,{\sc v} (top) and C\,{\sc iv} (bottom). Stellar lines are marked in
black, interstellar lines (i.s.) and IUE Resau marks (R) are marked in
green. }
\label{f_windlines_BD+37D1977}
\end{figure}

Finally we turn to the wind lines in the spectrum of
BD$+37^{\circ}1977$, which are the P\,Cygni profiles of the N\,{\sc v}
and the C\,{\sc iv} resonance doublets shown in the upper panel of
Fig.\,\ref{f_lines_BD+37D1977}, or in more detail in
Fig.\,\ref{f_windlines_BD+37D1977}. The terminal wind velocity
$v_\infty$ is adjusted to fit the width of the P\,Cygni profiles, while 
the mass-loss rate $\dot{M}$ influences their overall strength. The
parameters for the best fit are included in Table\,\ref{t_phys_new}. The
two wind-formed lines available are close to saturation. Hence the error
margin of the mass-loss rate (usually about a factor of two) is 
asymmetric in this case, i.e.\ much higher rates cannot be safely
excluded. One should also keep in mind that any error in the nitrogen and
carbon abundances directly propagates into our empirical mass-loss
rate. 

Note that in this semi-empirical approach, the mass-loss rate and wind
velocity are free parameters. Hydrodynamically consistent solutions that
agree with observations have been constructed for, e.g.\, massive O
stars \citep[for a recent review see][]{puls08}, and some Wolf-Rayet 
subtypes \citep{wr111,graefener08} under the paradigm of 
radiation-driven winds. Problems with such models are encountered not
only for very strong mass-loss, but also for very thin winds \citep[the
``weak wind problem'', cf.][]{marcolino09}. A first  indication if the
empirically found mass-loss is plausibly driven by radiation pressure is
given by the {\it work ratio}, defined as the total momentum intercepted
by the wind from the radiation field over the mechanical  momentum
carried by the stellar wind after escaping from the star
($\dot{M}v_{\infty}$). A computed work ratio somewhat below unity, say
about $0.5$, is expected because of the incompleteness of opacities in
the model calculations. Note however that for being fully
hydrodynamically consistent the momentum balance must be fulfilled not
only globally, but at each radial point.  In  the case of
BD$+37^{\circ}1977$, our final model has a work ratio of 0.60, which
means that the parameters are very plausible for a radiation-driven
wind.

\newcolumntype{.}{D{.}{.}{2.2}}
\begin{table*}
\caption{Current physical parameters for each program star. 
 Abundances $X$ are mass fraction in per cent. }
\label{t_phys_new}
\begin{tabular}{l c c c . . c . c c c c c}
\hline 
Star \rule[0mm]{0mm}{3.25mm} 
  & $T_{\rm eff}$ 
  & $E_{B-V}$  
  & $\log g$ 
  & \multicolumn{1}{c}{$\log L$} 
  & \multicolumn{1}{c}{$D\!M$} 
  & $v_{\infty}$ 
  & \multicolumn{1}{c}{$\log \dot{M}$} 
  & $X_{\rm H}$ 
  & $X_{\rm C}$ & $X_{\rm N}$ & $X_{\rm Si}$ & $X_{\rm Fe}$  \\
& [kK] 
  & [mag] 
  & [${\rm cm\,s^{-2}}$] 
  & \multicolumn{1}{c}{[${\rm L_\odot}$]} 
  & \multicolumn{1}{c}{[mag]} 
  & [${\rm km\,s^{-1}}$] 
  & \multicolumn{1}{c}{[${\rm M_{\odot}\,yr^{-1}}$]} 
  & [\%] &[\%] &[\%] &[\%] &[\%]      \\[1mm]
\hline 

BD$+37^{\circ}1977$ \rule[0mm]{0mm}{3.25mm} 
& 48.0 & 0.03 & 4.0$^\dagger$ & 4.4^\dagger & 12.2 & 2000 & -8.2  
& 0 & 2.5 & 0.3 & 0.08 & 0.06 \\[1mm]

BD$+37^{\circ}442$ 
& 48.0 & 0.09 & 4.0$^\dagger$ & 4.4^\dagger & 11.7 & 2000 & -8.5  
& 0$^\dagger$ & 2.5$^\dagger$ & 0.3$^\dagger$ & 0.08$^\dagger$ & 0.06 \\[1mm]
 
HD160641 
& 35.5 & 0.45 & 2.7$^\dagger$ & 4.5^\dagger & 11.8 & 500 & -7.3  
& 0 & 1.0 & 0.3 & 0.13 & 0.11  \\[1mm] 

BD$-9^{\circ}4395$
& 25.1 & 0.33 & 2.5$^\dagger$ & 4.4^\dagger & 13.4 & 400 & -7.9  &
 0.04$^\dagger$ & 1.3$^\dagger$ & 0.09$^\dagger$ & 0.14$^\dagger$ & 
 0.015$^\dagger$ \\[1mm] 

BD$+10^{\circ}2179$
& 18.5 & 0.06 & 2.6$^\dagger$ & 3.6^\dagger & 12.3 & 400 & -8.9 &
 0$^\dagger$ & 3.0$^\dagger$ & 0.08$^\dagger$ & 0.0005& 0.006 \\[1mm] 

HD144941
& 27.0 & 0.31 & 3.9$^\dagger$ & 2.7^\dagger &  8.4 & 500 & -9.8 &
 1.4$^\dagger$ &  0.005$^\dagger$ & 0.003$^\dagger$ & 
 0.001          & 0.01$^\dagger$ \\[1mm] 

Sun$^\ddagger$  & & & & & & & & &
0.21 & 0.06 & 0.066 & 0.11   \\

\hline 
\end{tabular}
\vspace{-2mm}
\flushleft
\begin{tabular}{l}
$^\dagger$ taken from (or at least oriented at) previous photospheric analyses 
          (cf.\ Table\,\ref{t_phys_old})\\
$^\ddagger$ Solar metal abundances from \citet{asplund05} 
          are given for comparison
\end{tabular}
\end{table*}

\subsection{BD+37$^{\circ}$442}

\begin{figure}
\includegraphics[width=\columnwidth]{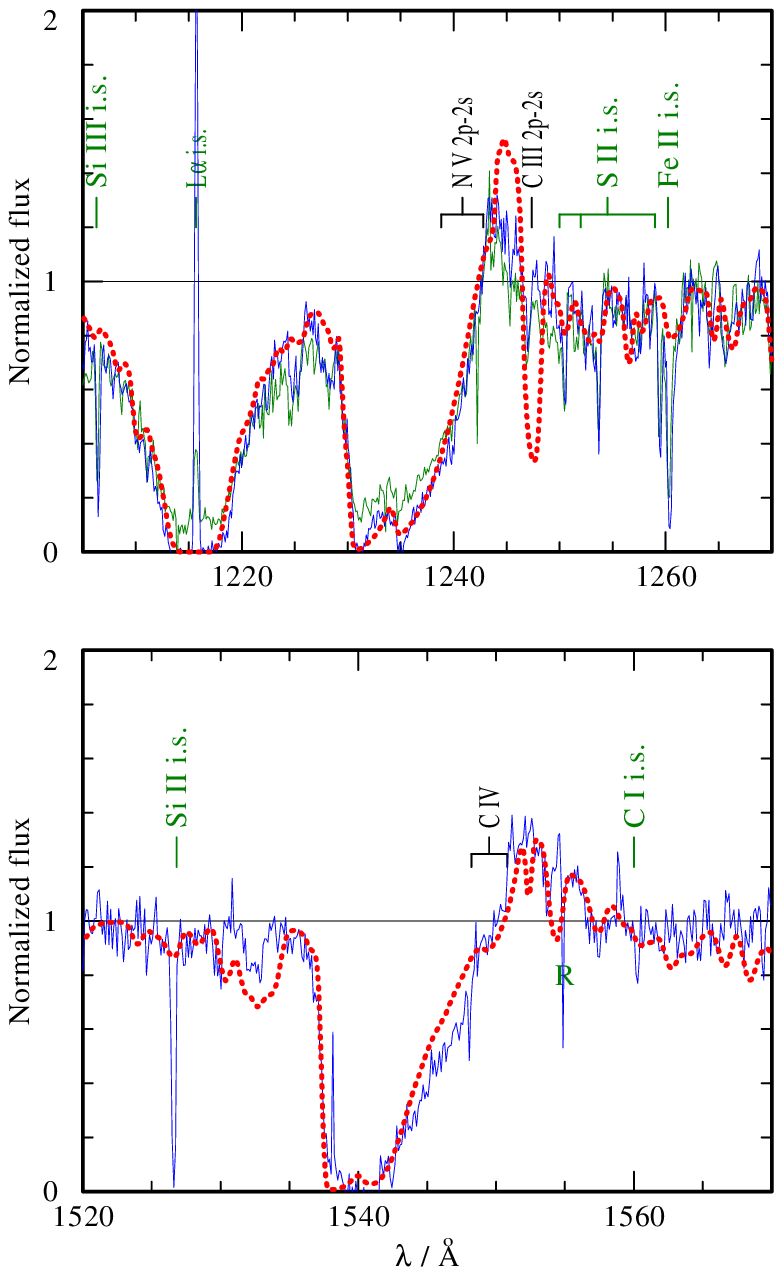}
\caption{Normalized line spectrum  for BD$+37^{\circ}442$ with 
the UV resonance doublets of
N\,{\sc v} (top) and C\,{\sc iv} (bottom). Observations are in blue and
green, while the best-fit model is red-dotted.}
\label{f_windlines_BD+37D442}
\end{figure}

BD$+37^{\circ}442$ is a spectroscopic twin of BD$+37^{\circ}1977$,
although previous analyses assigned a slightly higher effective
temperature to the former.  Within the uncertainties, we can adopt the
same best-fit model for both objects. BD$+37^{\circ}442$ shows more
interstellar redding, but still appears brighter than 
BD$+37^{\circ}1977$ which must be attributed to a smaller distance if
the intrinsic luminosities are the same (see Table\,\ref{t_phys_new}).
As mentioned above, we can take the chemical composition from a
photospheric analysis by \cite{bauer95}. The complete SED and line-fit
plots are provided as online material, while only the zoomed-in wind
lines are shown in the main paper (Fig.\,\ref{f_windlines_BD+37D442}).
The different observations for the N\,{\sc v} profile (blue: ORFEUS,
green: IUE) demonstrate the problems of background subtraction in 
high-resolution IUE data. 

\subsection{HD160641 (V2076 Oph)}

The helium absorption lines in the optical spectrum are well reproduced
with an effective temperature of 35.5\,kK, slightly higher than found in
previous work. The only abundance analysis published to date is a
``quick look-see'' by \citet{all54}. For the present study 
we set the carbon abundance to 1\,\% (mass fraction), which leads to a 
reasonable
reproduction of the photospheric C\,{\sc iii} and C\,{\sc iv} lines.
Such a carbon abundance of a few times solar seems to be typical for
extreme helium stars. For the nitrogen abundance the solar value also
produces too weak photospheric lines, while five times solar (0.3\% mass
fraction) gives a reasonable fit. The Si\,{\sc iv} line at 4089\,\AA\
forces us to adopt twice the solar silicon abundance. Solar iron
perfectly reproduces the UV iron forest, and hydrogen is apparently
absent. The iron forest has also significant impact on the SED, which is
perfectly fitted by the model. The fit of the SED and the UV and
blue-optical line spectrum are shown in the online material
(Fig.\,\ref{f_HD160641}). In the formal integral, different
microturbulent velocities are again used for the UV range with the wind lines
($v_{\rm D}=50\,{\rm km\,s^{-1}}$) and for the photospheric spectrum in
the optical ($25\,{\rm km\,s^{-1}}$).

\begin{figure}
\includegraphics[width=\columnwidth]{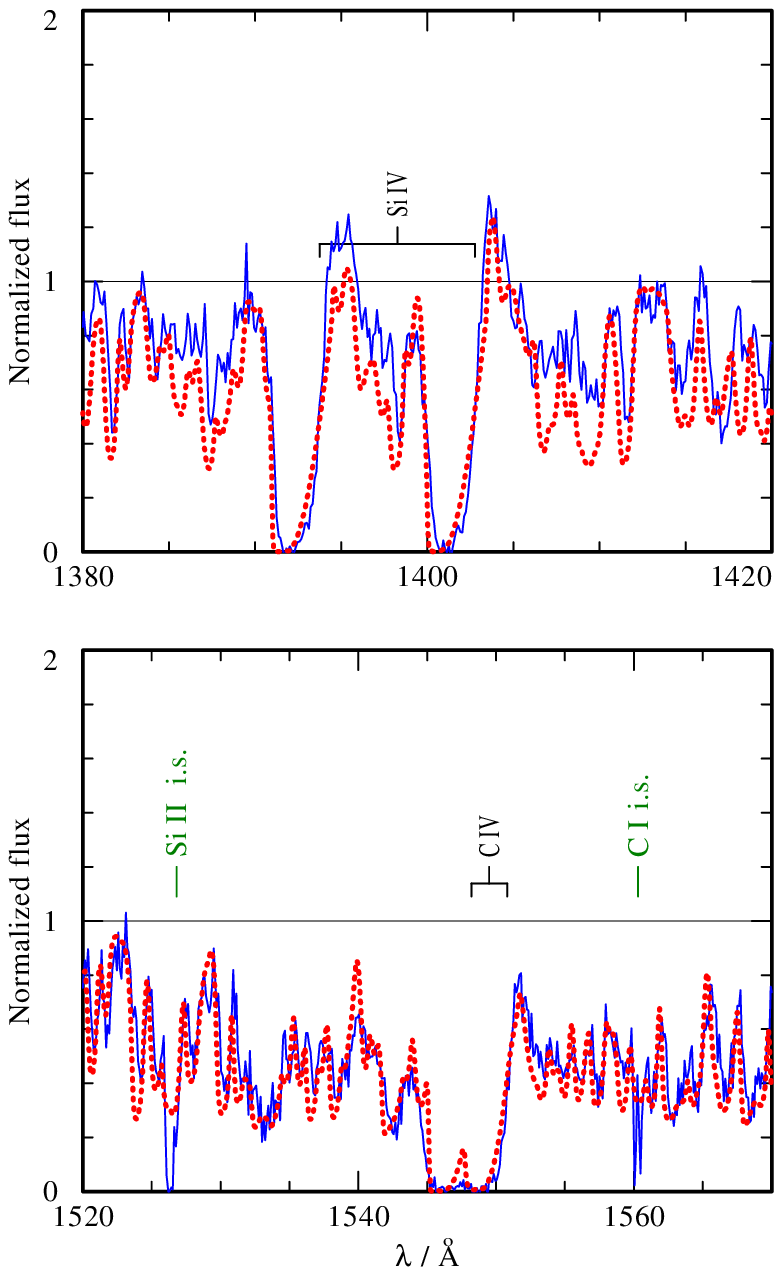}
\caption{Normalized line spectrum  for HD160641 with 
the UV resonance doublets of
Si\,{\sc iv} (top) and C\,{\sc iv} (bottom). Observations are in blue, 
while the best-fit model is red-dotted.}
\label{f_windlines_HD160641}
\end{figure}

The UV spectrum of HD160641 shows wind-line profiles of the Si\,{\sc iv}
and C\,{\sc iv} resonance doublets (see
Fig.\,\ref{f_windlines_HD160641}), both embedded in a pronounced forest
of iron lines. The P-Cygni profiles are not very wide, and fitted with a
moderate wind velocity  of $v_\infty = 500\,{\rm km\,s^{-1}}$. A
two-$\beta$-law, where 40\% of the velocity is attributed to a slow
acceleration with $\beta_2 = 4$, has improved the nice fit of the
profile shape.

By adjusting the mass-loss rate, the model matches both lines
simultaneously. Again, the fit is not very sensitive to higher values
because of saturation. Also for this star the ``work ratio'' is close to
unity (1.15) and supports the plausibility of a radiation-driven wind.   

The observation also shows the N{\sc v} resonance doublet as a P-Cygni
profile, while the model only predicts photospheric absorptions. 
Due to the relatively low stellar temperature, the ionization stage of 
N\,{\sc v} is not populated in the stellar wind. Hence the observation
of that wind line is a hint that ``superionization'', known from 
the winds of massive O stars, is also effective in the winds studied
here.

\begin{figure}
\includegraphics[width=\columnwidth]{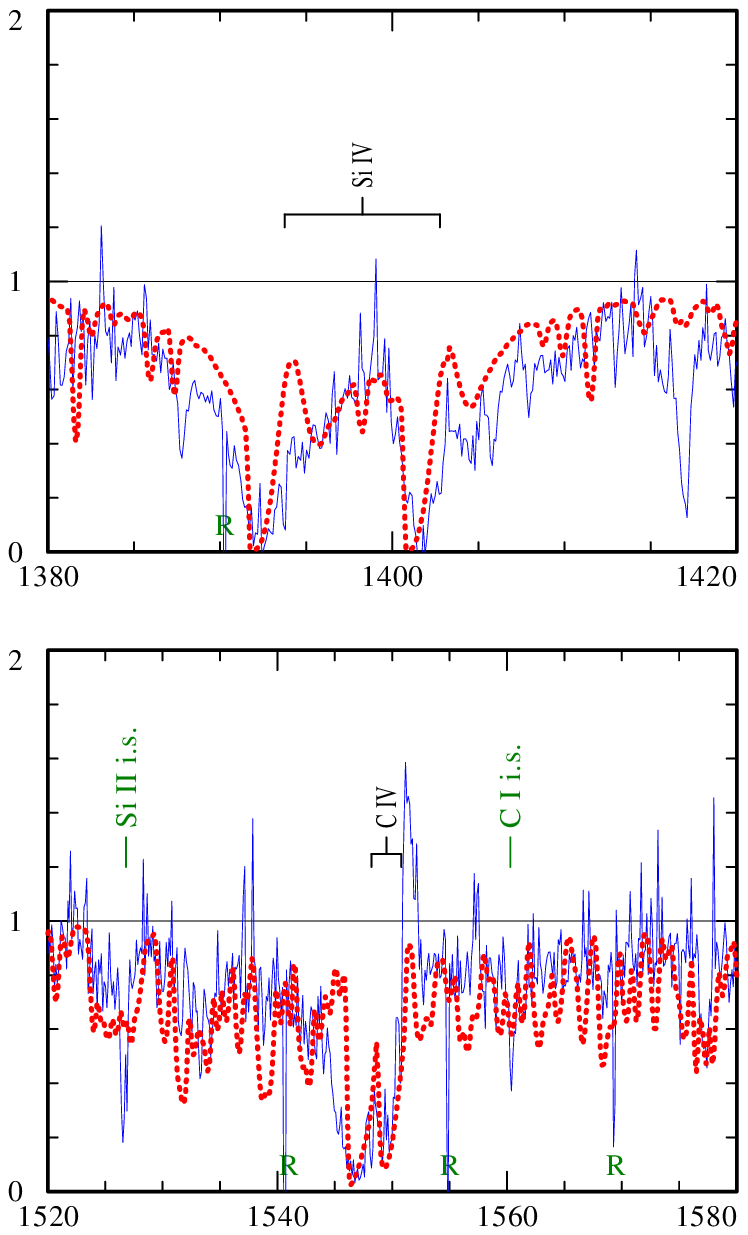}
\caption{Normalized line spectrum  for BD$-9^{\circ}4395$ with the UV
resonance doublets of Si\,{\sc iv} (top) and C\,{\sc iv} (bottom).
Observations are in blue, while the best-fit model is red-dotted.}
\label{f_windlines_BD-9D4395}
\end{figure}

\subsection{BD$-$9$^{\circ}$4395 (V2209\,Oph)}

BD$-9^{\circ}4395$ is a somewhat weird star, as it 
shows variations in total light, radial velocity and line spectrum.  
These peculiarities have been attributed to non-radial oscillations and 
dynamical activities in a circumstellar shell. 
A detailed analysis of this star has been published by \cite{jeffery92}. 

We adopt a slightly higher effective temperature (25.1\,kK) than this previous 
study (22.7\,kK) because otherwise our model predicts too strong lines 
from low ions like C\,{\sc ii}. Abundances have been determined in the 
previous photospheric analysis and can be adopted from there, leading to 
a good fit of the photospheric line spectrum (apart from the mentioned 
weird features) including the UV iron line forest (see Online Material,
Fig.\,\ref{f_BD-9D4395}). The SED fits perfectly when adopting the
reddening law from \cite{cardelli89} with $R_V=2.8$ (which gives
a marginally better match to the IUE data than the standard
$R_V=3.1$), and reveals a higher interstellar reddening than found 
in previous work.
 
The stellar wind manifests itself only marginally in the spectrum of 
BD$-9^{\circ}4395$. The C\,{\sc iv} resonance doublet forms a narrow 
P-Cygni profile, which can be fitted with a terminal wind velocity of 
$400\,{\rm km\,s^{-1}}$ and suitably adjusted mass-loss rate. However, 
even with a ``two-beta-law'' for the velocity field (40\% with $\beta_2 
= 4$), which worked fine for HD160641, the round shape of the absorption 
feature is not reproduced, and the calculated P-Cygni emission is weaker
than observed. Because of the small wind velocity, we calculated the
formal solution with a Doppler broadening of  $v_{\rm D}=30\,{\rm
km\,s^{-1}}$ in the UV and $15\,{\rm km\,s^{-1}}$ in the optical.

The Si\,{\sc iv} resonance doublet shows blue-shifted, asymmetric 
absorptions which are also not perfectly reproduced by the model. 
We attribute the remaining mismatch to the neglect of pressure broadening 
for resonance lines in the PoWR code. 

The final model for BD$-9^{\circ}4395$ has a ``work ratio'' of 1.06 and 
is thus nearly hydrodynamically consistent. 

\begin{figure}
\includegraphics[width=\columnwidth]{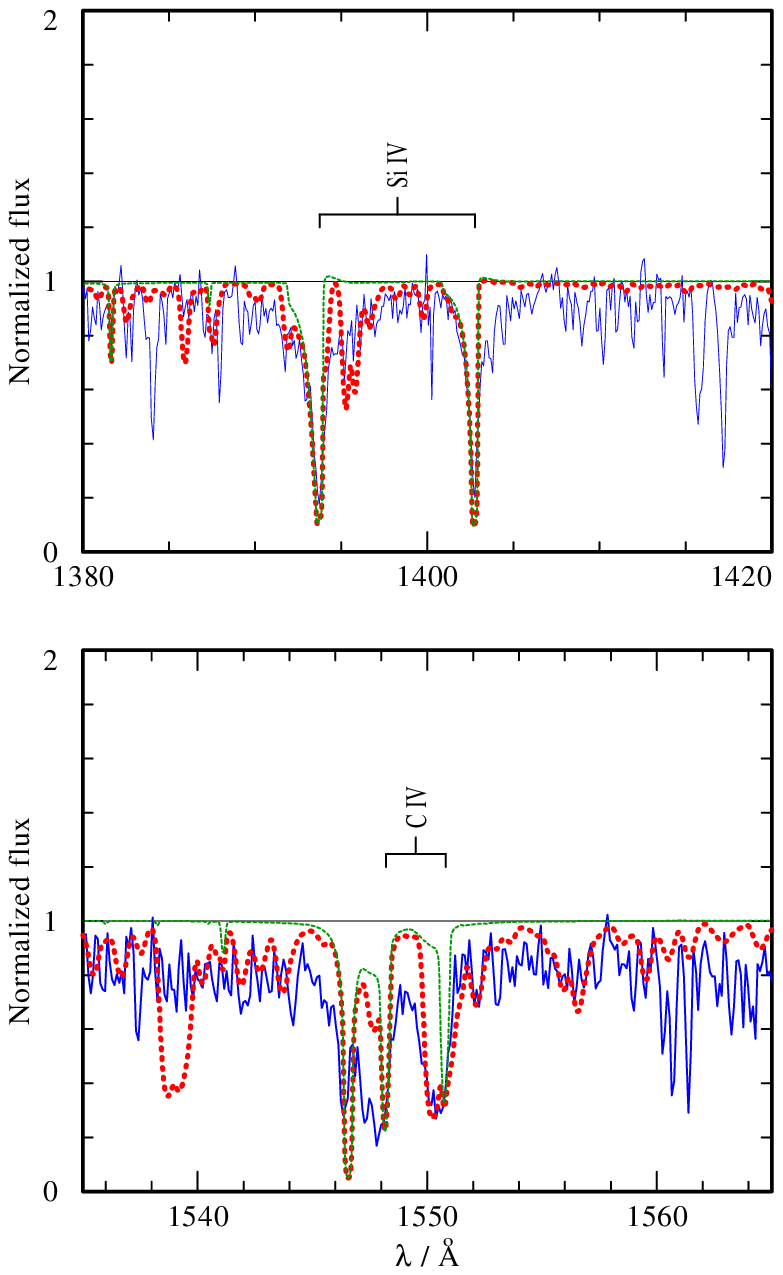}
\caption{Normalized line spectrum for BD$+10^{\circ}2179$ with  the UV
resonance doublets of Si\,{\sc iv} and C\,{\sc iv}. Observation is in
blue, while the best-fit model is red-dotted. The thin-dashed green
profiles are calculated without iron opacities in the formal integral.}
\label{f_windlines_BD+10D2179}
\end{figure}

\subsection{BD+10$^{\circ}$2179}

\label{sect_BD+10D2179}

This star is the coolest of our sample. Slightly discrepant values have
been published for its effective temperature (19.5 and 16.9\,kK,
respectively, cf.\ Table\,\ref{t_phys_old}). Therefore we inspect the
flux distribution   and the photospheric line spectrum. The SED fit for
models with  $T_{\ast} < 18$\,kK implies $E_{B-V}=0.00$, which is
unlikely  and in conflict with the interstellar Ly$\alpha$ absorption.
However, $T_{\ast} > 18$\,kK makes C\,{\sc ii} line at 1335\,\AA\ weaker
than observed. Only with $T_{\ast} = 19$\,kK, the Stark-broadened He{\sc
i}  line wings fit nicely (keeping $\log g=2.6$ from previous
work). As a
compromise we finally adopt $T_{\rm eff} = 18.5\,{\rm kK}$. 

The element abundances also differ between the two previous photospheric
analyses. Starting with the values from \cite{pandey06}, we found that
silicon must be even less abundant to avoid too strong photospheric
absorption features of the Si\,{\sc iv} resonance doublet. For carbon we
prefer the slightly higher abundance from \cite{heber83}.   

Only weak indications of a stellar wind are found in the UV spectrum   
of BD$+10^{\circ}2179$. The Si\,{\sc iv} resonance doublet shows
blue-shifted, asymmetric absorptions. When adopting a two-beta velocity field
with $v_\infty = 400\,{\rm km s^{-1}}$ and a suitable mass-loss
rate, the model can roughly reproduce these profiles
(Fig.\,\ref{f_windlines_BD+10D2179}). But since the wind profile is not
fully developed, $v_\infty$ is not accurately constrained. 
Because of the weak wind, we calculated all non-iron lines in the
formal solution with a small microturbulence of $15\,{\rm km\,s^{-1}}$.

From the C\,{\sc iv} resonance doublet, only the red component seems to
be reproduced by the same model. However, the feature apparently fitting
the red C\,{\sc iv} component is in fact due to iron. This is
demonstrated in Fig.\,\ref{f_windlines_BD+10D2179} by the thin, green
dashed profile which is calculated with the iron forest switched off. Thus
we must state that the model does not reproduce the C\,{\sc iv} resonance
doublet as a wind line at all. We found no escape from this discrepancy.
A (much) larger mass-loss rate makes not only the Si\,{\sc iv} resonance
doublet too strong, but also transforms the C\,{\sc ii} line at 1335\,\AA\
into a wind-line profile which is not observed. Reason for the missing 
C\,{\sc iv} wind line is that models for such low $T_{\rm eff}$ predict
that carbon recombines to C\,{\sc iii}  and C\,{\sc ii}. We must
conclude that also in this star some ``superionization''
keeps the ionization higher than predicted by our stationary models.     
As we will discuss below (\ref{sect_discussion}), frictional heating in 
this very thin wind may provide a viable mechanism. 

Our ``final model'' and the mass-loss rate thus relies only on the 
Si\,{\sc iv} resonance doublet, and should be taken with care. 
The ``work ratio'' is 0.41, i.e.\ not far from what is expected for a
radiation-driven wind. 

\begin{figure}
\includegraphics[width=\columnwidth]{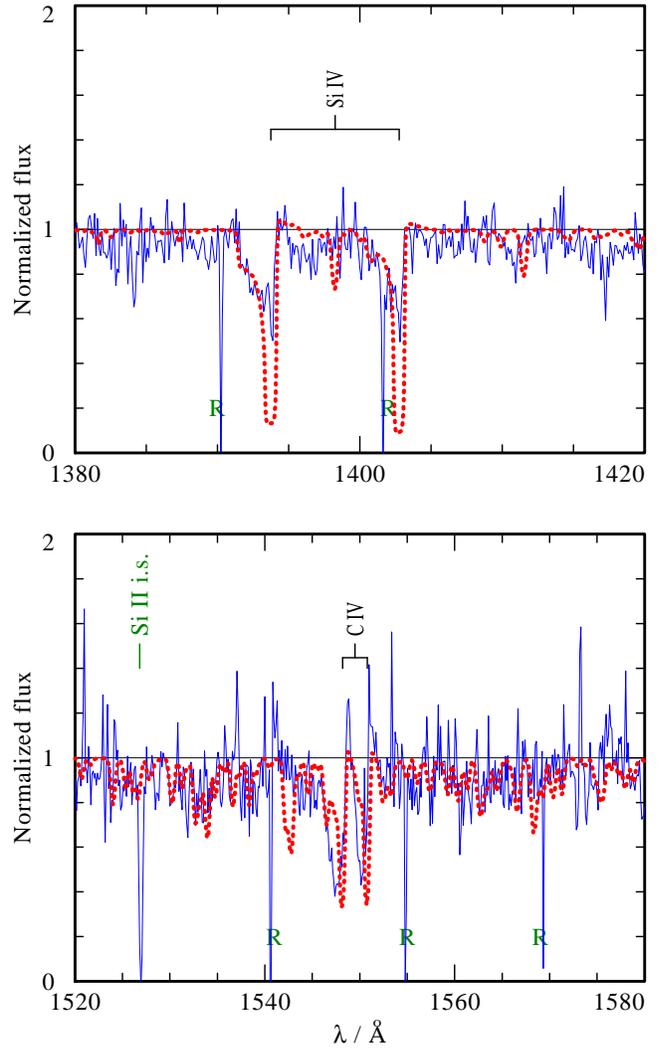}
\caption{Normalized line spectrum for HD144941 with 
the UV resonance doublets of
Si\,{\sc iv} (top) and C\,{\sc iv} (bottom). Observations are in blue, 
while the best-fit model is red-dotted.}
\label{f_windlines_HD144941}
\end{figure}

\subsection{HD144941}

Previous photospheric analyses give quite discrepant effective
temperatures for this star (see Table\,\ref{t_phys_old}). From a
detailed analysis of the optical line spectrum, \cite{harrison97} 
derived $T_{\rm eff} = 23.2\,{\rm kK}$. In contrast, \cite{jeffery01a}
obtained 27.8\,kK from fitting the UV flux distribution. We adopt 
$T_{\rm eff} = 27.0\,{\rm kK}$ as a compromise, but take  the gravity
and derived luminosity from \cite{harrison97}. Some of the 
pressure-broadened He\,{\sc i} lines appear too broad when compared to
our model (see Fig.\,\ref{f_HD144941} in the Online Material). 

The chemical abundances were also determined from the photospheric
spectrum in previous work \citep{harrison97}. Comparison with our model
(see Fig.\,\ref{f_HD144941} in the Online Material) gives the impression
that the traces of hydrogen might have been under-estimated. The
formal integral is calculated for the non-iron lines with a
microturbulence of  $v_{\rm D}=30\,{\rm km\,s^{-1}}$ in the UV and
$15\,{\rm km\,s^{-1}}$ in the optical.

The spectral signatures of the stellar wind are seen in the UV resonance
doublets of Si\,{\sc iv} and C\,{\sc iv} (see
Fig.\,\ref{f_windlines_HD144941}). In both cases, there are blueward
extended absorption features accompanied by weak, red-shifted P-Cygni
emissions. The reproduction of the wind profiles requires a terminal
wind speed of about $500\,{\rm km s^{-1}}$. The mass-loss rate of the
final model (see Table\,\ref{t_phys_new}) is a compromise, leaving the
wind feature of the C\,{\sc iv} doublet too weak. 

For both resonance doublets, the model shows too strong photospheric
absorption features at the rest wavelengths shining through the wind. In
the case of Si\,{\sc iv} this discrepancy was so strong that we felt the
need to reduce the silicon abundance against the  value from the
literature (from 25 to 10 ppm mass fraction), but the unshifted features
are still stronger than observed. 

We also tested the lower effective temperature (23.2\,kK) from
\cite{harrison97} for the wind-line fit. With these models, the problem
with the too strong photospheric components almost disappears. However, 
$T_{\rm eff} = 23.2\,{\rm kK}$ is too low to produce the observed
the C\,{\sc iv} wind line in our radiative-equilibrium models. 
In this case, one would need to argue that C\,{\sc iv} is due to some
``superionization'', as in the case of N\,{\sc v} in HD160641 and
C\,{\sc iv} in BD$+10^{\circ}2179$ (see above). 

These ambiguities make the derived mass-loss rate especially
uncertain. The work ratio for the final model is only 0.2,
possibly indicating that the mass-loss rate is over- or the luminosity
under-estimated.

\subsection{Wind variability}

%
%
\begin{figure*}
\includegraphics[width=12cm,angle=0]{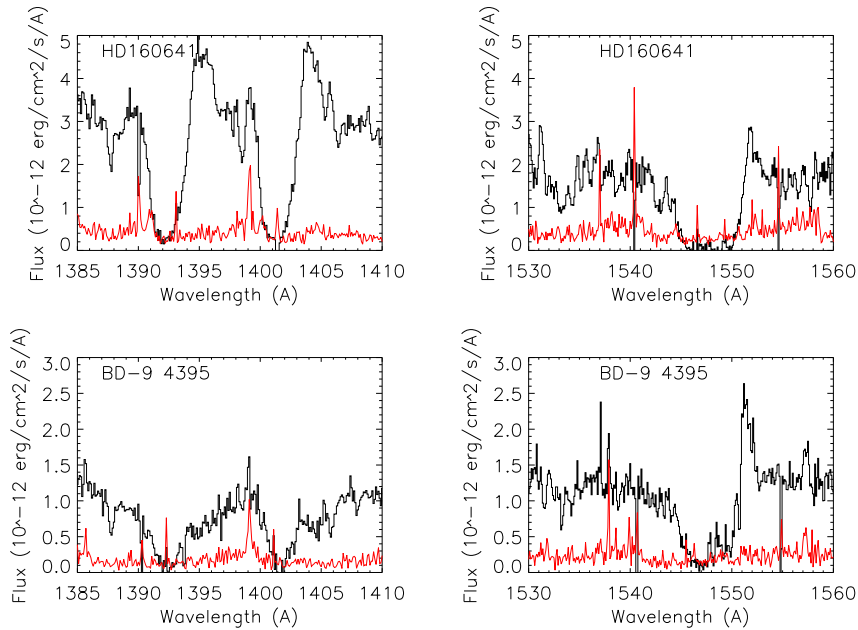}
\caption[UV LPV]{The Si\,{\sc iv} and C\,{\sc iv}
  ultraviolet resonance doublets for two extreme helium stars known to
  be intrinsic variables as observed with IUE. 
 The black histogram shows the mean obtained by combining all IUE
 HIRES and HST STIS images at these wavelengths. The red polyline
 shows the standard deviation about this mean.  }
\label{f_lpv_uv}
\end{figure*}

Many extreme helium stars are known to be intrinsically variable in both
light and radial velocity on timescales of one to several days
\citep{lynasgray87,jeffery92}. Repeat high-resolution spectra of
HD160641 and BD$-9^{\circ}4395$ were obtained with IUE on timescales of
days and years. \citet{jeffery92} found no evidence for variability of
the wind lines in BD$-9^{\circ}4395$. Figure~\ref{f_lpv_uv} shows the
mean  profile of  two resonance lines with strong wind profiles in boths
stars. It also shows the standard deviation of the individual
observations around the mean. While there is a hint of increased scatter
in the blue wing of some lines, it is no greater than seen in other
nearby regions of spectrum. It is concluded that there is little firm
evidence for any variation in the wind line profiles. In contrast,
photospheric lines are known to show show strong profile variability
\citep{jeffery92,wright06}.

\section{Conclusion}

\label{sect_discussion}

\begin{figure}
\includegraphics[width=\columnwidth]{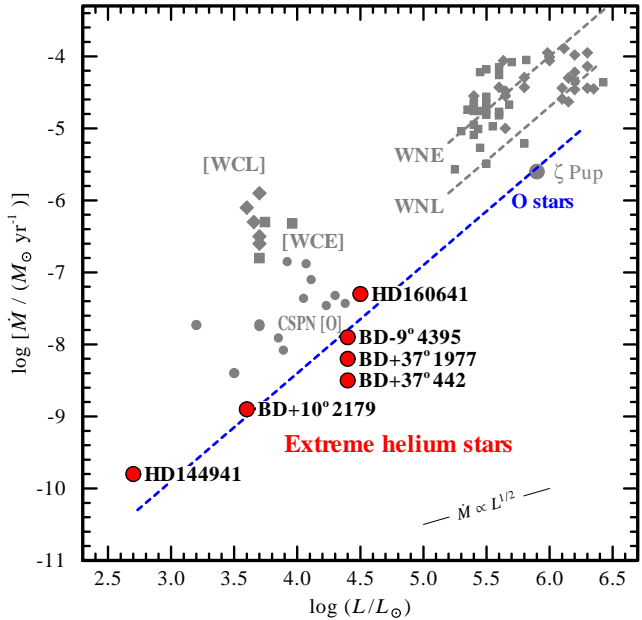}
\caption{Mass-loss rate versus luminosity for the studied extreme helium
stars (labelled red symbols), compared to various other groups of hot
stars from high to low masses. Grey groups of symbols denote massive
Wolf-Rayet stars (squares: WNE, diamonds: WNL). Also in grey are 
central stars of  planetary nebulae of hydrogen-rich composition
(CSPN\,[O], dots) or of Wolf-Rayet type (squares: [WCE], diamonds:
[WCL]). The long blue dashed line gives the slope of $\dot{M} \propto
L^{1.5}$, tight to the protopype O-type supergiant $\zeta$\,Puppis.}
\label{f_mdot-l}
\end{figure}

The spectra of six extreme helium stars under investigation are found to
indicate weak mass loss. This evidence comes from resonance
lines in the UV, which show P\,Cygni profiles or at least asymmetric
absoption features. The hotter two of our programme stars, with $T_{\rm
eff} > 45\,{\rm kK}$, exhibit wind profiles of the N\,{\sc iv} and
C\,{\sc iv} resonance doublet, while the cooler stars ($T_{\rm
eff} < 35\,{\rm kK}$) show these effects in Si\,{\sc iv} and
C\,{\sc iv}. 

By fitting these wind lines to model spectra, we have determined the
mass-loss rates. The result depends inversely on the abundance of the
respective element, which we have mostly taken from previous analyses of
the photospheric spectra. Thus one must realize that, 
apart from the uncertainties of the fit, a possible error in the element
abundance directly translates into a wrong mass-loss rate. 

One should also be aware that the empirical mass-loss rate scales with
the adopted distance to the star. The reason is that  the resonance-line
fits basically measure an optical depth which, in a homologuous scaling,
depends on the density times the linear dimension. The density depends on
$\dot{M}/R^2$, and therefore the optical depth   scales as 
$\dot{M}/R_\ast$. For models with the same optical depth, we thus obtain
$\dot{M} \propto R$ and, since $L \propto R^2$,   a scaling  $\dot{M}
\propto L^{1/2}$.  

Moreover, the fit relies on the ionization structure of the models
calculated in radiative equilibrium. When the ion under consideration is
abundant, this should be reliable. However, in some of the cooler stars
carbon is mainly in the stages C\,{\sc ii} and  C\,{\sc iii}. In such
cases the traces of the observed  C\,{\sc iv} ion can be largely
enhanced by additional ionization processes, such as that caused by
hydrodynamic shocks in the stellar wind which emit radiation in the
hard-UV and X-ray range. Such ``superionization'' is well known from the
winds of massive hot stars. Mass-loss rates cannot be reliably derived
from ``superionized'' lines.   

Thus we warn that the mass-loss rates should be taken with care,
although the error margin is impossible to quantify for the individual
stars. 

As a general result, stellar winds appear as a persistent feature of hot
stars, irrespective of their wide range of luminosities,
masses and chemical composition. In Fig.\,\ref{f_mdot-l} we plot the
empirical mass-loss rates $\dot{M}$ over the stellar luminosity $L$,
both on a logarithmic scale. The theory of radiation-driven winds in its
most elementary form \citep{CAK} predicts a correlation of about   
$\dot{M} \propto L^{1.5}$, which roughly remains valid in later
refinements of this prediction such as the ``modified wind momentum --
luminosity relationship'' \citep[see e.g.][]{puls08}. Fig.\,\ref{f_mdot-l}
reveals that this relation, represented by the blue dashed line
normalized to the values for the massive O-type supergiant
$\zeta$\,Puppis, indeed marks a kind of lower limit for the mass loss
from hot stars of all kinds.

Our extreme helium stars lie close to this relation. Among the massive
stars, the Wolf-Rayet types show considerably stronger mass loss.
Similarly high $\dot{M}$ are found for the early subtypes (WNE) and the late
subtypes (WNL), but the latter ones have on the average higher
luminosities.  

Mass loss is also found from central stars of planetary nebulae (CSPN),
which are hot stars of lower mass and less luminosity than the massive
early-type stars. In their majority the CSPN are hydrogen rich
(CSPN-[O]). Their mass-loss rates lie about one order of magnitude above
the dashed $\dot{M}$-$L$-relation in Fig.\,\ref{f_mdot-l}. Extremely
strong is the mass loss from the hydrogen deficient CSPN with Wolf-Rayet
type spectra. According to detailed hydrodynamic modelling, the extremely
strong mass loss from Wolf-Rayet stars is caused by their proximity to
the Eddington limit, and driven by multiple line scattering
\citep{graefener08}. 

The proximity to the Eddington limit can be measured by the ratio
between the inward force by gravitation and the outward force by
radiation pressure, $\Gamma = g_{\rm rad} / g_{\rm grav}$.
Only accounting for the radiation pressure on free electrons by Thomson
scattering, the Eddington $\Gamma$ follows from the effective temperature 
and the surface gravity as 
\begin{equation}
\Gamma = 10^{-15.12}\  q\  T_{\rm eff}^4 / g_{\rm grav}
\end{equation}
where $T_{\rm eff}$ is in Kelvin and $g$ in cgs-units. For our stars we
may set the number of free electrons per atomic mass unit to $q=0.5$ for
fully ionized helium.

To evaluate $\Gamma$, we must realize that the
gravities $g$ given in Table\,\ref{t_phys_new} are spectroscopically
determined from the pressure-broadening of lines, and thus are {\em
effective} gravities $g = g_{\rm grav} (1-\Gamma)$. Taking this into
account, we obtain the $\Gamma$ values compiled in Table\,\ref{t_gamma}. 
The correlation with the mass-loss rates is striking, especially within 
the group of four stars with nearly the same luminosity. The outstandingly 
high mass loss from HD160641 is thus a consequence of its low gravity.

\begin{table}
\caption{Ratio between the inward force by gravitation and the outward 
force by radiation pressure, $\Gamma = g_{\rm rad} / g_{\rm grav}$.}
\label{t_gamma}
\centering
\begin{tabular}{lc}
\hline 
Star & $\Gamma$ \\
\hline 
HD160641                & 0.55  
\rule[0mm]{0mm}{3.25mm}  \\
BD$-9^{\circ}4395$      & 0.32 \\
BD$+37^{\circ}1977$     & 0.17 \\
BD$+37^{\circ}442$      & 0.17 \\
BD$+10^{\circ}2179$     & 0.10 \\
HD144941                & 0.03 \\
\hline 
\end{tabular}
\end{table}

This agreement with the expectations for radiation driven winds, also 
reflected by the plausible ``work ratios'' for the individual stars as 
mentioned in Sect.\,\ref{Results}, may be considered 
an independent argument that the luminosities and distances are correct at 
least within an order of magnitude. Adopting a different luminosity
would shift the empirical mass-loss rate parallel to the 
thin line with slope 1/2 indicated in the lower-right 
corner of Fig.\,\ref{f_mdot-l}. 

It has been theoretically predicted that in very thin winds the
radiation pressure will accelerate only the metal ions, while the bulk
matter of helium stays inert. This {\em ion decoupling} will lead to
frictional heating, or perhaps even to instabilities that completely
disrupt the smooth stellar wind \citep{springmann92,krticka01}. Among
our programme stars, the thinnest wind is encountered at
BD$+10{^\circ}2179$. According to our final model, the number density at
one stellar radius from the photosphere is about $10^{8.0}$ atoms per
cubic centimeter. The mass-loss rate and radius of that star are almost
identical to those of $\tau$\,Sco, for which ion decoupling has been
predicted. Hence we may speculate that the ``superionization'' invoked
for C\,{\sc iv} (Sect.\,\ref{sect_BD+10D2179}) is caused by frictional
heating from ion decoupling. The same may also hold for HD144941, the
star with the second-thinnest wind ($\log n_{\rm atom}/{\rm cm}^{-3} = 8.4$
at $r=2\,R_\ast$). The consistent wind-line fits for the other programme
stars provide evidence that ion decoupling does not take place in 
their winds.

\section*{Acknowledgments}

This publication makes use of data products from the Two Micron All
Sky Survey, which is a joint project of the University of
Massachusetts and the Infrared Processing and Analysis
Center/California Institute of Technology, funded by the National
Aeronautics and Space Administration and the National Science
Foundation.

Some of the data presented in this paper were obtained from the
Multimission Archive at the Space Telescope Science Institute
(MAST). STScI is operated by the Association of Universities for
Research in Astronomy, Inc., under NASA contract NAS5-26555. Support
for MAST for non-HST data is provided by the NASA Office of Space
Science via grant NAG5-7584 and by other grants and contracts.

Based on INES data from the IUE satellite.

The Armagh Observatory is funded by direct grant from the Northern
Ireland Dept of Culture Arts and Leisure.

\bibliographystyle{mn2e}
\bibliography{mnemonic,ehe}

\begin{thebibliography}{}

\bibitem[\protect\citeauthoryear{{Aller}}{{Aller}}{1954}]{all54}
{Aller} L.~H.,  1954, Memoires of the Societe Royale des Sciences de Liege, 1,
  337

\bibitem[\protect\citeauthoryear{{Asplund}, {Grevesse} \& {Sauval}}{{Asplund}
  et~al.}{2005}]{asplund05}
{Asplund} M.,  {Grevesse} N.,    {Sauval} A.~J.,  2005, in {Barnes} III T.~G.,
  {Bash} F.~N.,  eds, Cosmic Abundances as Records of Stellar Evolution and
  Nucleosynthesis Vol.~336 of Astronomical Society of the Pacific Conference
  Series, {The Solar Chemical Composition}.
p.~25

\bibitem[\protect\citeauthoryear{{Bartolini}, {Bonifazi}, {Fusi Pecci},
  {Oculi}, {Piccioni}, {Serra} \& {Dantona}}{{Bartolini}
  et~al.}{1982}]{bartolini82}
{Bartolini} C.,  {Bonifazi} A.,  {Fusi Pecci} F.,  {Oculi} L.,  {Piccioni} A.,
  {Serra} R.,    {Dantona} F.,  1982, Ap\&SS, 83, 287

\bibitem[\protect\citeauthoryear{{Bauer} \& {Husfeld}}{{Bauer} \&
  {Husfeld}}{1995}]{bauer95}
{Bauer} F.,  {Husfeld} D.,  1995, A\&A, 300, 481

\bibitem[\protect\citeauthoryear{{Cardelli}, {Clayton} \& {Mathis}}{{Cardelli}
  et~al.}{1989}]{cardelli89}
{Cardelli} J.~A.,  {Clayton} G.~C.,    {Mathis} J.~S.,  1989, \apj, 345, 245

\bibitem[\protect\citeauthoryear{{Castor}, {Abbott} \& {Klein}}{{Castor}
  et~al.}{1975}]{CAK}
{Castor} J.~I.,  {Abbott} D.~C.,    {Klein} R.~I.,  1975, \apj, 195, 157

\bibitem[\protect\citeauthoryear{{Clayton}, {Geballe}, {Herwig}, {Fryer} \&
  {Asplund}}{{Clayton} et~al.}{2007}]{clayton07}
{Clayton} G.~C.,  {Geballe} T.~R.,  {Herwig} F.,  {Fryer} C.,    {Asplund} M.,
  2007, ApJ, 662, 1220

\bibitem[\protect\citeauthoryear{{Cowley}}{{Cowley}}{1971}]{Cowley}
{Cowley} C.~R.,  1971, The Observatory, 91, 139

\bibitem[\protect\citeauthoryear{{Darius}, {Giddings} \& {Wilson}}{{Darius}
  et~al.}{1979}]{darius79}
{Darius} J.,  {Giddings} J.~R.,    {Wilson} R.,  1979, in {Willis} A.~J.,  ed.,
  The first year of IUE; Proceedings of the Symposium, London, England, April
  4-6, 1979. (A80-16301 04-90) London, University College {Discovery of mass
  loss from hot subdwarfs}.
pp 363--370

\bibitem[\protect\citeauthoryear{{Dudley} \& {Jeffery}}{{Dudley} \&
  {Jeffery}}{1992}]{dudley92}
{Dudley} R.~E.,  {Jeffery} C.~S.,  1992, in {Heber} U.,  {Jeffery} C.~S.,  eds,
  The Atmospheres of Early-Type Stars Vol.~401 of Lecture Notes in Physics,
  Berlin Springer Verlag, {Mass loss from upsilon Sgr and other helium stars}.
p.~298

\bibitem[\protect\citeauthoryear{{Giddings}}{{Giddings}}{1981}]{giddings81}
{Giddings} J.~R.,  1981, PhD thesis, University of London

\bibitem[\protect\citeauthoryear{{Gr{\"a}fener} \& {Hamann}}{{Gr{\"a}fener} \&
  {Hamann}}{2005}]{wr111}
{Gr{\"a}fener} G.,  {Hamann} W.-R.,  2005, \aap, 432, 633

\bibitem[\protect\citeauthoryear{{Gr{\"a}fener} \& {Hamann}}{{Gr{\"a}fener} \&
  {Hamann}}{2008}]{graefener08}
{Gr{\"a}fener} G.,  {Hamann} W.-R.,  2008, \aap, 482, 945

\bibitem[\protect\citeauthoryear{{Griem}, {Baranger}, {Kolb} \&
  {Oertel}}{{Griem} et~al.}{1962}]{Griem}
{Griem} H.~R.,  {Baranger} M.,  {Kolb} A.~C.,    {Oertel} G.,  1962, Physical
  Review, 125, 177

\bibitem[\protect\citeauthoryear{{Hamann}, {Feldmeier} \& {Oskinova}}{{Hamann}
  et~al.}{2008}]{clumpingworkshop}
{Hamann} W.-R.,  {Feldmeier} A.,    {Oskinova} L.~M.,  eds, 2008, {Clumping in
  hot-star winds}

\bibitem[\protect\citeauthoryear{{Hamann} \& {Gr{\"a}fener}}{{Hamann} \&
  {Gr{\"a}fener}}{2004}]{PoWR}
{Hamann} W.-R.,  {Gr{\"a}fener} G.,  2004, \aap, 427, 697

\bibitem[\protect\citeauthoryear{{Hamann}, {Schoenberner} \& {Heber}}{{Hamann}
  et~al.}{1982}]{hamann82}
{Hamann} W.-R.,  {Schoenberner} D.,    {Heber} U.,  1982, A\&A, 116, 273

\bibitem[\protect\citeauthoryear{{Harrison} \& {Jeffery}}{{Harrison} \&
  {Jeffery}}{1997}]{harrison97}
{Harrison} P.~M.,  {Jeffery} C.~S.,  1997, A\&A, 323, 177

\bibitem[\protect\citeauthoryear{{Heber}}{{Heber}}{1983}]{heber83}
{Heber} U.,  1983, A\&A, 118, 39

\bibitem[\protect\citeauthoryear{{Husfeld}}{{Husfeld}}{1987}]{husfeld87}
{Husfeld} D.,  1987, in {Philip} A.~G.~D.,  {Hayes} D.~S.,   {Liebert} J.~W.,
  eds, IAU Colloq. 95: Second Conference on Faint Blue Stars {Non-LTE analyses
  of extremely helium-rich subluminous O stars}.
pp 237--246

\bibitem[\protect\citeauthoryear{{Iben} Jr. \& {Tutukov}}{{Iben} \&
  {Tutukov}}{1984}]{iben84}
{Iben} Jr. I.,  {Tutukov} A.~V.,  1984, ApJS, 54, 335

\bibitem[\protect\citeauthoryear{{Jeffery}}{{Jeffery}}{1988}]{jeffery88}
{Jeffery} C.~S.,  1988, \mnras, 235, 1287

\bibitem[\protect\citeauthoryear{{Jeffery}}{{Jeffery}}{2008}]{jeffery08b}
{Jeffery} C.~S.,  2008, in {Werner} A.,  {Rauch} T.,  eds, Hydrogen-Deficient
  Stars Vol.~391 of Astronomical Society of the Pacific Conference Series,
  {Hydrogen-Deficient Stars: An Introduction}.
p.~3

\bibitem[\protect\citeauthoryear{{Jeffery} \& {Harrison}}{{Jeffery} \&
  {Harrison}}{1997}]{jeffery97}
{Jeffery} C.~S.,  {Harrison} P.~M.,  1997, A\&A, 323, 393

\bibitem[\protect\citeauthoryear{{Jeffery} \& {Heber}}{{Jeffery} \&
  {Heber}}{1992}]{jeffery92}
{Jeffery} C.~S.,  {Heber} U.,  1992, A\&A, 260, 133

\bibitem[\protect\citeauthoryear{{Jeffery}, {Heber} \& {Hamann}}{{Jeffery}
  et~al.}{1986}]{jeffery86}
{Jeffery} C.~S.,  {Heber} U.,    {Hamann} W.~R.,  1986, in {Rolfe} E.~J.,  ed.,
  New Insights in Astrophysics. Eight Years of UV Astronomy with IUE Vol.~263
  of ESA Special Publication, {Ultraviolet spectroscopy of the
  hydrogen-deficient star HD 144941: The energy distribution and the stellar
  wind}.
pp 369--372

\bibitem[\protect\citeauthoryear{{Jeffery}, {Starling}, {Hill} \&
  {Pollacco}}{{Jeffery} et~al.}{2001}]{jeffery01a}
{Jeffery} C.~S.,  {Starling} R.~L.~C.,  {Hill} P.~W.,    {Pollacco} D.,  2001,
  MNRAS, 321, 111

\bibitem[\protect\citeauthoryear{{Kondo}, {Boggess} \& {Maran}}{{Kondo}
  et~al.}{1989}]{kondo89}
{Kondo} Y.,  {Boggess} A.,    {Maran} S.~P.,  1989, ARA\&A, 27, 397

\bibitem[\protect\citeauthoryear{{Krti{\v c}ka} \& {Kub{\'a}t}}{{Krti{\v c}ka}
  \& {Kub{\'a}t}}{2001}]{krticka01}
{Krti{\v c}ka} J.,  {Kub{\'a}t} J.,  2001, \aap, 369, 222

\bibitem[\protect\citeauthoryear{{Kudritzki}, {Puls}, {Lennon}, {Venn},
  {Reetz}, {Najarro}, {McCarthy} \& {Herrero}}{{Kudritzki}
  et~al.}{1999}]{kudritzki99}
{Kudritzki} R.~P.,  {Puls} J.,  {Lennon} D.~J.,  {Venn} K.~A.,  {Reetz} J.,
  {Najarro} F.,  {McCarthy} J.~K.,    {Herrero} A.,  1999, A\&A, 350, 970

\bibitem[\protect\citeauthoryear{{Lynas-Gray}, {Kilkenny}, {Skillen} \&
  {Jeffery}}{{Lynas-Gray} et~al.}{1987}]{lynasgray87}
{Lynas-Gray} A.~E.,  {Kilkenny} D.,  {Skillen} I.,    {Jeffery} C.~S.,  1987,
  MNRAS, 227, 1073

\bibitem[\protect\citeauthoryear{{Marcolino}, {Bouret}, {Martins}, {Hillier},
  {Lanz} \& {Escolano}}{{Marcolino} et~al.}{2009}]{marcolino09}
{Marcolino} W.~L.~F.,  {Bouret} J.-C.,  {Martins} F.,  {Hillier} D.~J.,  {Lanz}
  T.,    {Escolano} C.,  2009, \aap, 498, 837

\bibitem[\protect\citeauthoryear{{Oskinova}, {Hamann} \&
  {Feldmeier}}{{Oskinova} et~al.}{2007}]{macroclumping}
{Oskinova} L.~M.,  {Hamann} W.-R.,    {Feldmeier} A.,  2007, \aap, 476, 1331

\bibitem[\protect\citeauthoryear{{Pandey}, {Lambert}, {Jeffery} \&
  {Rao}}{{Pandey} et~al.}{2006}]{pandey06}
{Pandey} G.,  {Lambert} D.~L.,  {Jeffery} C.~S.,    {Rao} N.~K.,  2006, ApJ,
  638, 454

\bibitem[\protect\citeauthoryear{{Puls}, {Vink} \& {Najarro}}{{Puls}
  et~al.}{2008}]{puls08}
{Puls} J.,  {Vink} J.~S.,    {Najarro} F.,  2008, \aapr, 16, 209

\bibitem[\protect\citeauthoryear{{Rauch}}{{Rauch}}{1996}]{rauch96}
{Rauch} T.,  1996, in {C.~S.~Jeffery \& U.~Heber} ed., Hydrogen Deficient Stars
  Vol.~96 of Astronomical Society of the Pacific Conference Series, {NLTE
  analysis of the extreme helium star HD 160641}.
pp 174--+

\bibitem[\protect\citeauthoryear{{Saio} \& {Jeffery}}{{Saio} \&
  {Jeffery}}{1988}]{saio88b}
{Saio} H.,  {Jeffery} C.~S.,  1988, ApJ, 328, 714

\bibitem[\protect\citeauthoryear{{Saio} \& {Jeffery}}{{Saio} \&
  {Jeffery}}{2002}]{saio02}
{Saio} H.,  {Jeffery} C.~S.,  2002, MNRAS, 333, 121

\bibitem[\protect\citeauthoryear{{Sch\"{o}nberner}}{{Sch\"{o}nberner}}{1986}]{%
schoenberner86}
{Sch\"{o}nberner} D.,  1986, in {Hunger} K.,  {Schoenberner} D.,   {Kameswara
  Rao} N.,  eds, IAU Colloq. 87: Hydrogen Deficient Stars and Related Objects
  Vol.~128 of Astrophysics and Space Science Library, {Evolutionary status and
  origin of extremely hydrogen-deficient stars}.
pp 471--480

\bibitem[\protect\citeauthoryear{{Sch\"oning} \& {Butler}}{{Sch\"oning} \&
  {Butler}}{1989a}]{sch89c}
{Sch\"oning} T.,  {Butler} K.,  1989a, A\&AS, 79, 153

\bibitem[\protect\citeauthoryear{{Sch\"oning} \& {Butler}}{{Sch\"oning} \&
  {Butler}}{1989b}]{sch89b}
{Sch\"oning} T.,  {Butler} K.,  1989b, A\&AS, 78, 51

\bibitem[\protect\citeauthoryear{{Sch\"oning} \& {Butler}}{{Sch\"oning} \&
  {Butler}}{1989c}]{sch89a}
{Sch\"oning} T.,  {Butler} K.,  1989c, A\&A, 219, 326

\bibitem[\protect\citeauthoryear{{Springmann} \& {Pauldrach}}{{Springmann} \&
  {Pauldrach}}{1992}]{springmann92}
{Springmann} U.~W.~E.,  {Pauldrach} A.~W.~A.,  1992, \aap, 262, 515

\bibitem[\protect\citeauthoryear{{Vidal}, {Cooper} \& {Smith}}{{Vidal}
  et~al.}{1973}]{VCS}
{Vidal} C.~R.,  {Cooper} J.,    {Smith} E.~W.,  1973, \apjs, 25, 37

\bibitem[\protect\citeauthoryear{{Webbink}}{{Webbink}}{1984}]{webbink84}
{Webbink} R.~F.,  1984, ApJ, 277, 355

\bibitem[\protect\citeauthoryear{{Wolff}, {Pilachowski} \&
  {Wolstencroft}}{{Wolff} et~al.}{1974}]{wolff74}
{Wolff} S.~C.,  {Pilachowski} C.~A.,    {Wolstencroft} R.~D.,  1974, ApJ, 194,
  L83

\bibitem[\protect\citeauthoryear{{Wright}, {Lynas-Gray}, {Kilkenny},
  {Cottrell}, {Shobbrook}, {Koen}, {van Wyk}, {Kilmartin}, {Martinez} \&
  {Gilmore}}{{Wright} et~al.}{2006}]{wright06}
{Wright} D.~J.,  {Lynas-Gray} A.~E.,  {Kilkenny} D.,  {Cottrell} P.~L.,
  {Shobbrook} R.~R.,  {Koen} C.,  {van Wyk} F.~W.,  {Kilmartin} P.~M.,
  {Martinez} P.,    {Gilmore} A.~C.,  2006, MNRAS, 369, 2049

\end{thebibliography}

\label{lastpage}


\clearpage

\appendix{\Large\bf Online Material}

\newpage

\begin{figure*}
\includegraphics[width=\textwidth,angle=0]{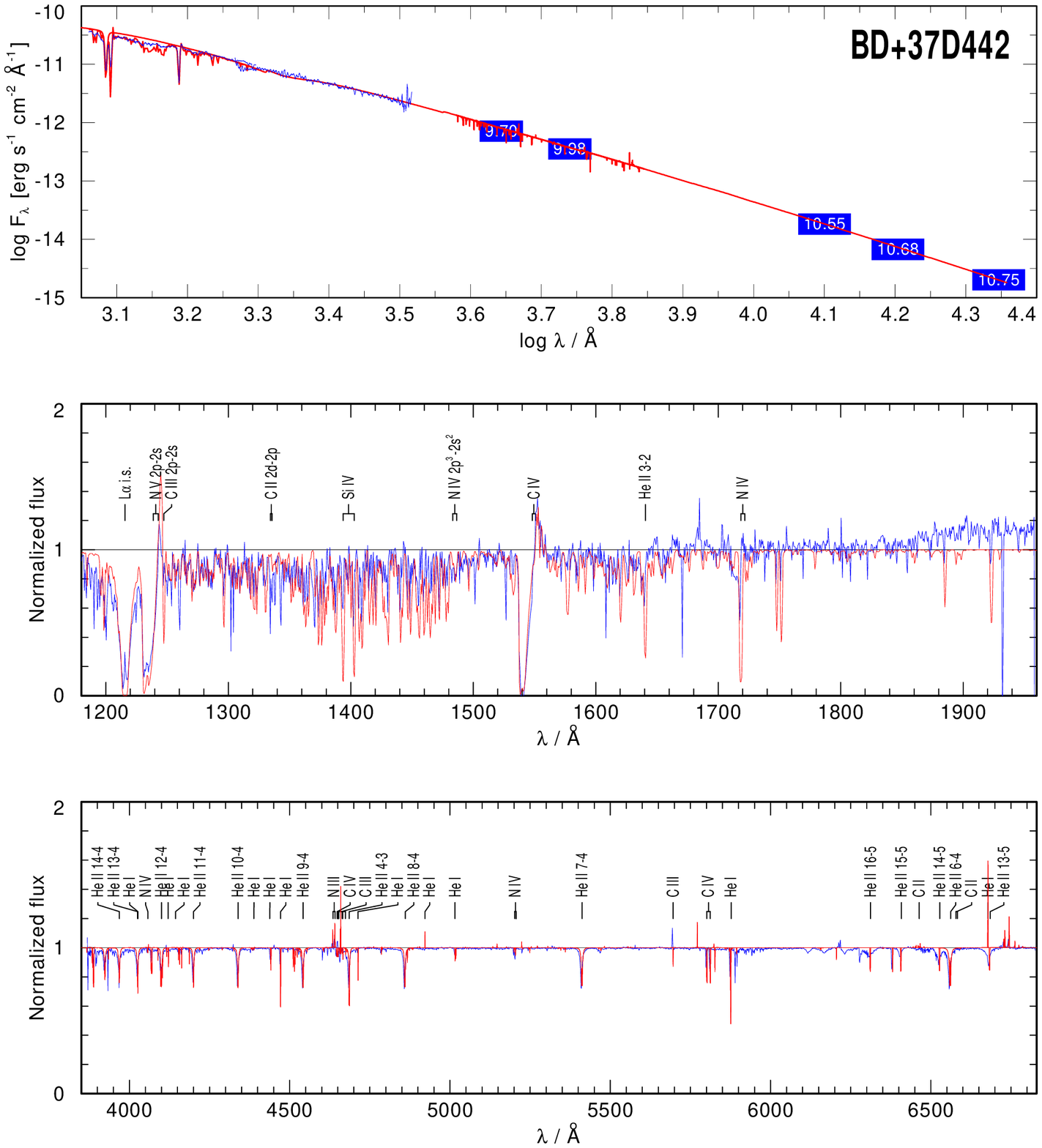}
\caption{BD$+37{^\circ}442$. {\em Top panel:}  Spectral energy
distribution (SED) for  from the UV to the near IR. The observations
comprise  low-resolution IUE data (thin blue line), visual photometry
(blue blocks  with magnitudes) and near-IR photometry from 2MASS. The
model continuum (see Table\,\ref{t_phys_new} for the parameters) is
shown  by the straight red line, in parts of the spectrum augmented by
the  synthetic line spectrum. {\em Middle panel:} Line spectrum  in the
UV; the observation (blue line) has been divided by the model continuum
for normalization. The model spectrum is shown by the red-dotted line. 
{\em Bottom panel:} Same for the optical range. This observation has
been normalized to the continuum  ``by eye''.}
\label{f_BD+37D442}
\end{figure*}

\begin{figure*}
\includegraphics[width=\textwidth,angle=0]{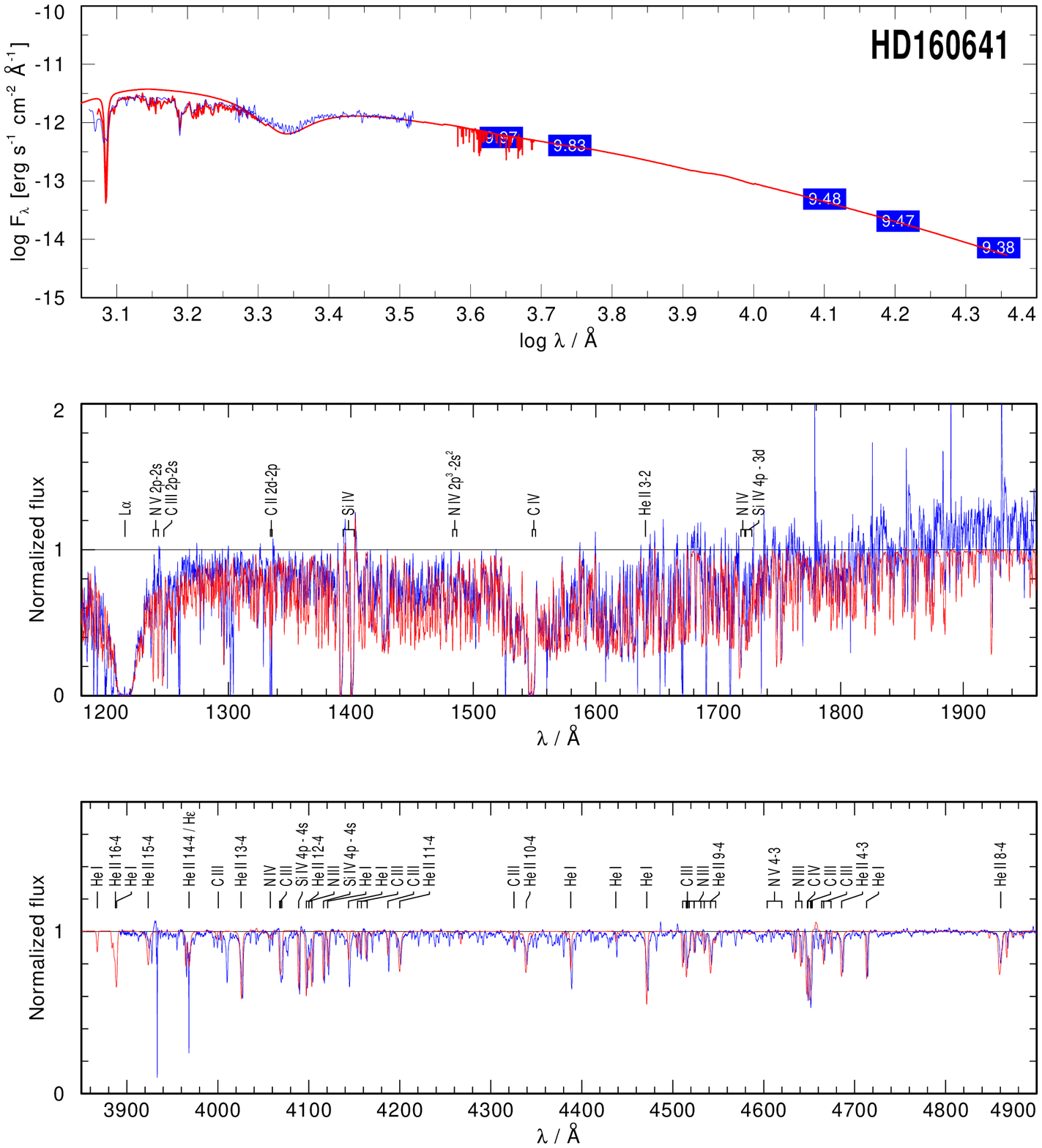}
\caption{Same as previous figure, but for HD160641}
\label{f_HD160641}
\end{figure*}

\begin{figure*}
\includegraphics[width=\textwidth,angle=0]{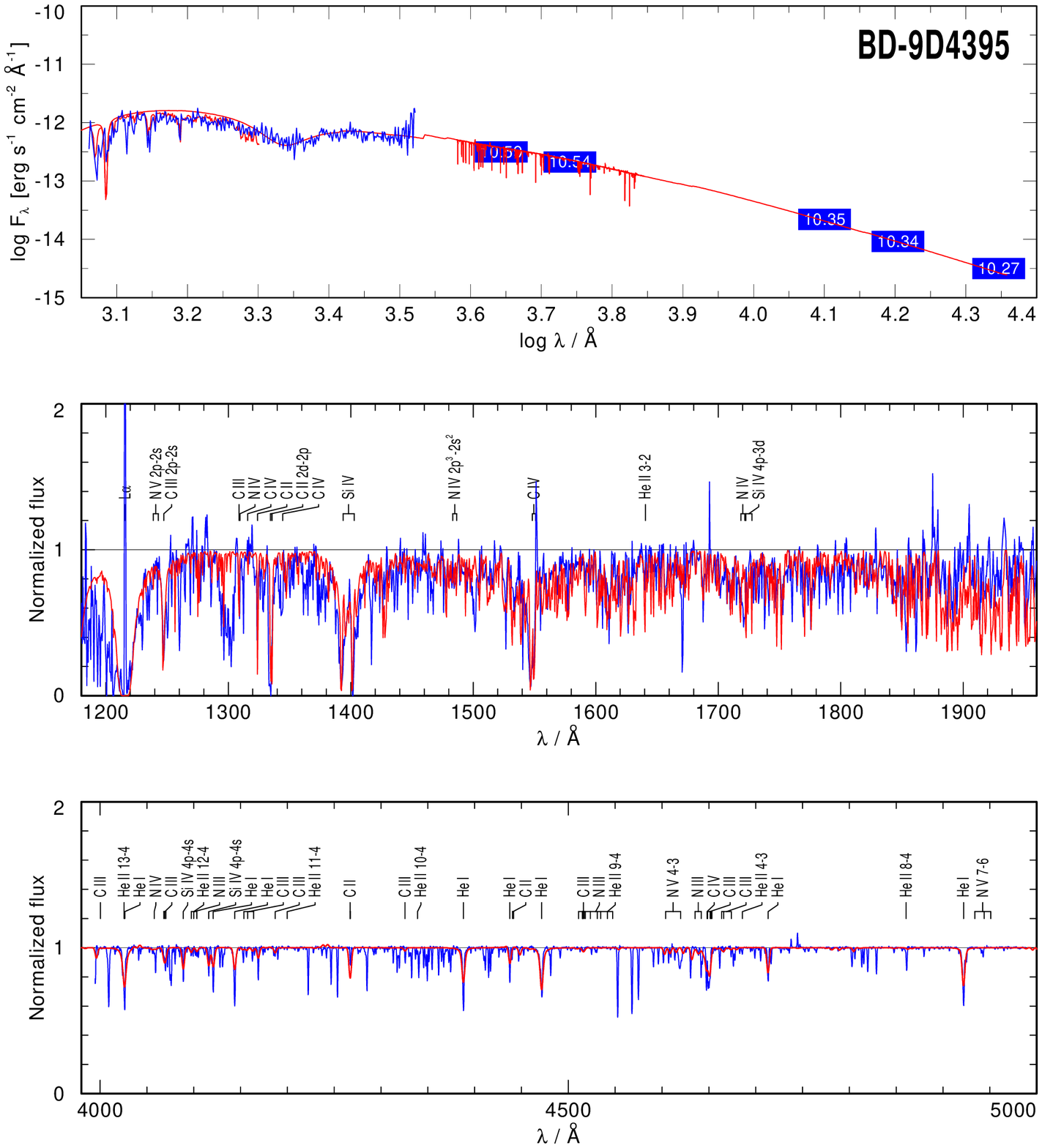}
\caption{Same as previous figure, but for BD$-9^{\circ}4395$}
\label{f_BD-9D4395}
\end{figure*}

\begin{figure*}
\includegraphics[width=\textwidth,angle=0]{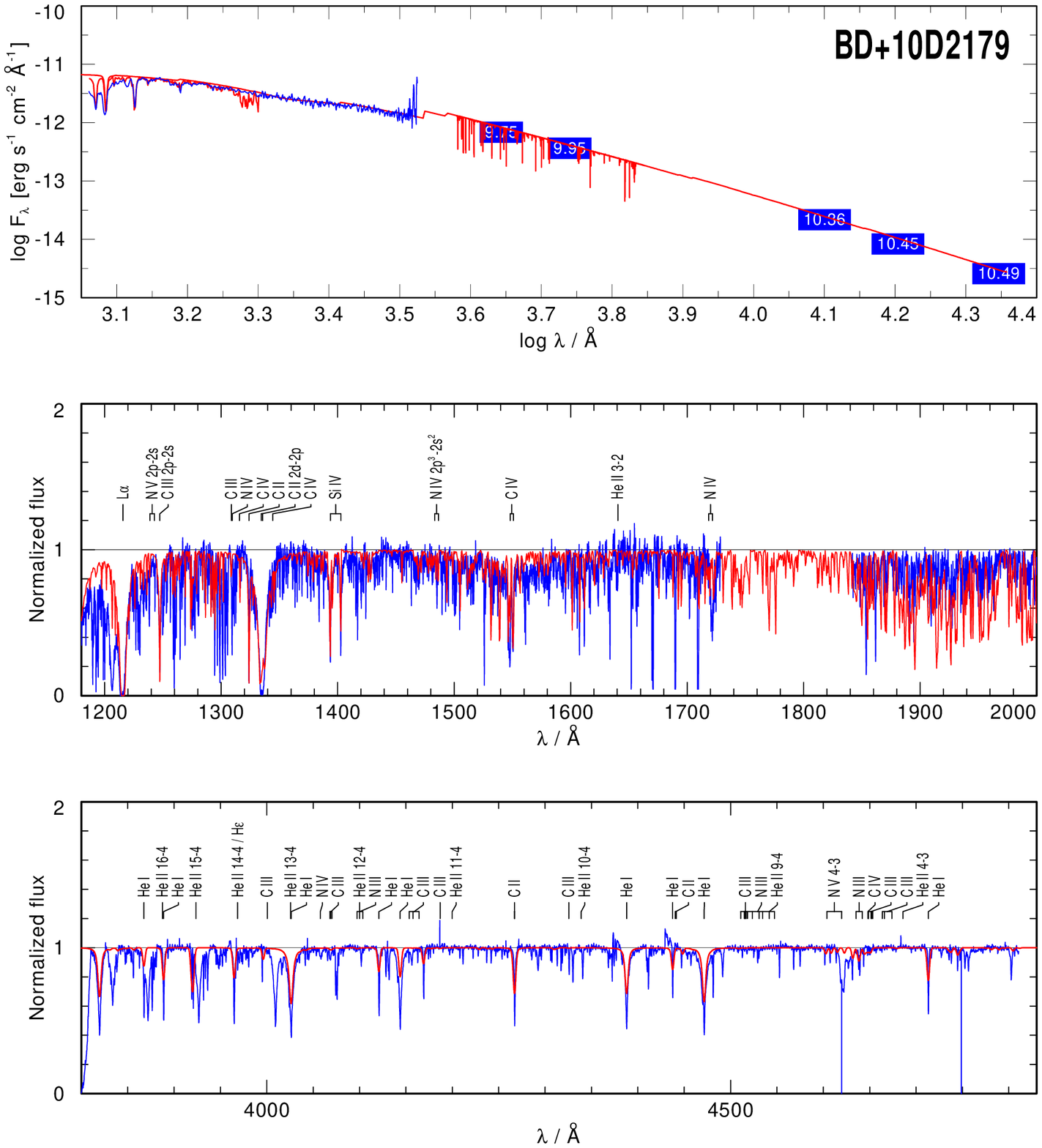}
\caption{Same as previous figure, but for BD$+10^{\circ}2179$}
\label{f_BD+10D2179}
\end{figure*}

\begin{figure*}
\includegraphics[width=\textwidth,angle=0]{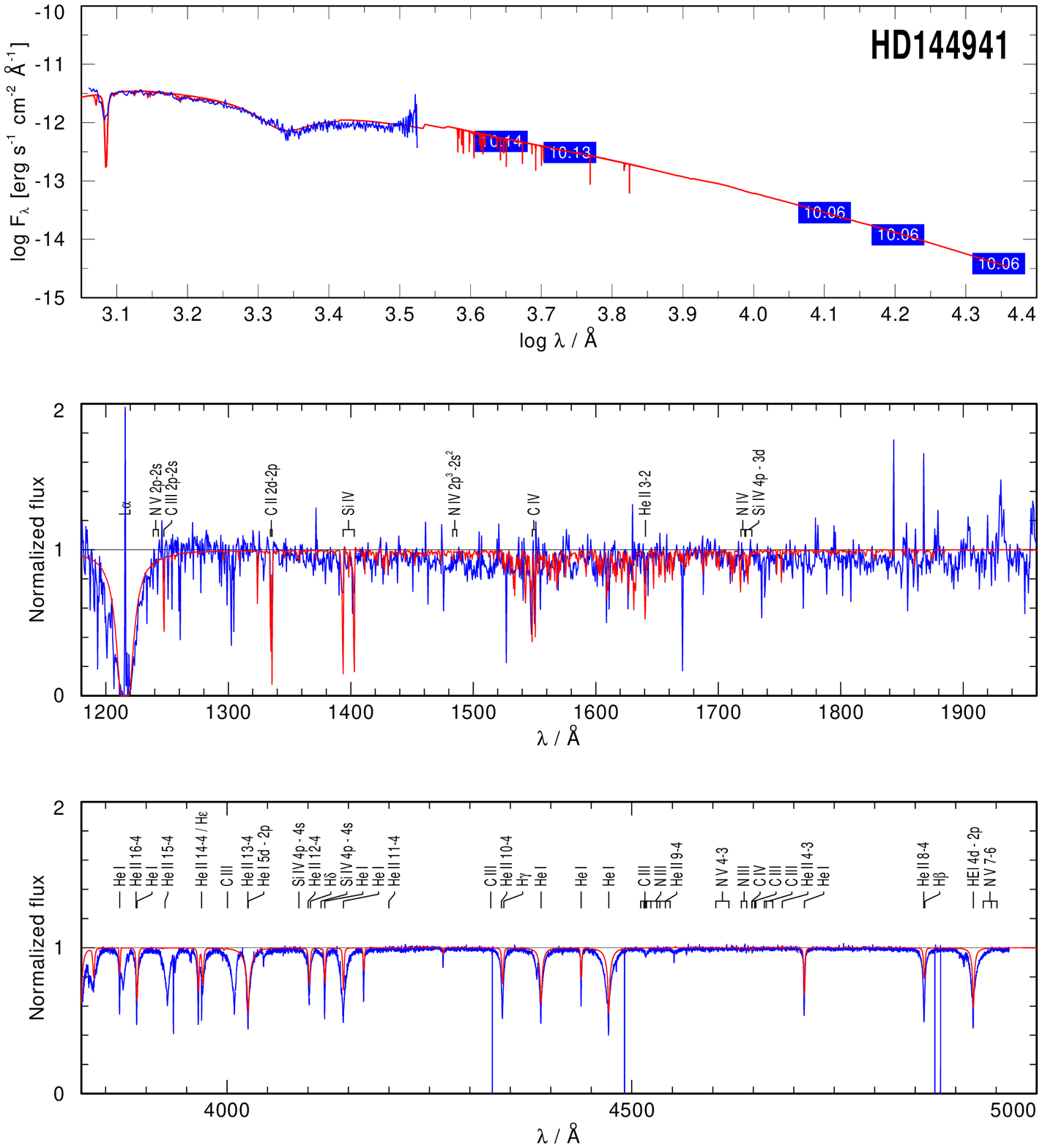}
\caption{Same as previous figure, but for HD144941}
\label{f_HD144941}
\end{figure*}

\end{document}